\documentclass[11pt]{article}
\usepackage{a4wide}
\usepackage{amsmath,amssymb,amsfonts}
\usepackage{comment}
\usepackage{epsfig,verbatim,graphics}
\usepackage{slashed,siunitx}
\usepackage[table,x11names]{xcolor}
\usepackage{cite}
\usepackage{caption}
\usepackage{subcaption}

\usepackage{hyperref}

\newcommand{\mathsym}[1]{{}}

\newcommand{\eref}[1]{(\ref{#1})}

\renewcommand\({\left(}
\renewcommand\){\right)}
\renewcommand\[{\left[}
\renewcommand\]{\right]}

\newcommand{\dd}{{\rm d}}
\declareslashed{}{{^-}}{.1}{0.1}{\rm d}

\newcommand{\e}{{\rm e}}
\newcommand{\is}{{\hat =}}
\newcommand\vp{\varphi}
\newcommand\eps{\epsilon}
\newcommand\mpl{m_{\rm P}}
\newcommand\GeV{{\rm GeV}}

\newcommand\wi{\omega_{k,0}}

\def\ba{\begin{eqnarray}}
\def\ea{\end{eqnarray}}
\def\be{\begin{equation}}
\def\ee{\end{equation}}

\def\A{\mathcal{A}}
\def\L{\mathcal{L}}
\def\O{\mathcal{O}}

\def\W{\mathcal{W}}

\def\del{\nabla}
\def\nn{\nonumber}
\def\({\left(}
\def\){\right)}

\def\eref#1{(\ref{#1})}

\newcommand{\roughly}[1]{\mathrel{\raise.3ex\hbox{$#1$\kern-0.85em
\lower1ex\hbox{$\sim$}}}}

\begin{document}
\begin{titlepage}
\begin{center}

\hfill Nikhef 2017-010

\vskip 1.5cm

{\LARGE \bf Electroweak stability and non-minimal coupling}

\vskip 1cm

\renewcommand\star{\thefootnote}{\fnsymbol{footnote}}
\setcounter{footnote}{0}

{\bf
Marieke Postma\footnote{{\tt mpostma@nikhef.nl },
        \footnotemark[2]{\tt jorindev@nikhef.nl}} and
   Jorinde van de Vis\footnotemark[2]
}

\renewcommand*{\thefootnote}{\number{footnote}}
\setcounter{footnote}{0}

\vskip 25pt

{\em 
Nikhef, \\Science Park 105, \\1098 XG Amsterdam, The Netherlands
}

\end{center}

\vskip 0.5cm

\begin{center} {\bf ABSTRACT}\\[3ex]\end{center} 

The measured values of the Higgs and top quark mass indicate that the
electroweak vacuum is metastable if there is no new physics below the
Planck scale. This is at odds with a period of high scale inflation. A
non-minimal coupling between the Higgs field and the Ricci scalar can
stabilize the vacuum as it generates a large effective Higgs mass
during inflation. We consider the effect of this coupling during
preheating, when Higgs modes can be produced very efficiently due to
the oscillating Ricci scalar. We compute their effect on the effective
potential and the energy density.  The Higgs excitations are defined with
respect to the adiabatic vacuum. We study the adiabaticity conditions
and find that the dependence of our results on the choice of the order
of the adiabatic vacuum increases with time. For large enough coupling
particle production is so efficient that the Higgs decays to the true
vacuum before this is an issue. However, for smaller values of the
Higgs-curvature coupling no definite statements can be made as the
vacuum dependence is large.

\end{titlepage}

\newpage
\setcounter{page}{1} \tableofcontents

\newpage


\section{Introduction}

No evidence for physics beyond the Standard Model (SM) has been found
so far at the Large Hadron Collider or any other experiment.  This
opens up the possibility of the desert scenario, in which the SM is
valid up to Planck scale energies. If true, our universe might be
metastable. Indeed, with the particle content of the SM, the Higgs
quartic coupling $\lambda$ runs to negative values at large energy
scales, and the Higgs potential develops a second minimum in addition
to the electroweak minimum.  For the best fit values of the top quark
and Higgs mass the quartic Higgs coupling becomes zero
$\lambda(\mu_{\rm cr}) =0$ at a scale $\mu_{\rm cr}\sim 10^{11}$ GeV,
although stability of the SM up to the Planck scale is only excluded
at the 2-3$\sigma$ level \cite{disc1, disc2, branchina,branchina2,
  archil, alexss,kniehl}. In this paper we will take the instability
of the Higgs potential at face value, and consider its cosmological
consequences.

The presence of the global minimum at large field values poses no
danger at present as the metastable vacuum can only decay via a slow
tunnelling process, and has a lifetime that is much longer than the age
of the universe.  Within a cosmological context the longevity of our
vacuum is no longer assured. If the Hubble scale $H$ during inflation
is comparable or larger than the maximum of the Higgs potential, then
vacuum decay is no longer suppressed as the Higgs field can quantum
fluctuate over the potential barrier
\cite{Espinosa:2007qp,Enqvist:2014bua,Fairbairn:2014zia,Kobakhidze:2013tn,Hook:2014uia}.
This is a concern for large field inflationary models such as chaotic
inflation and Starobinsky type inflation. It was pointed out that
stability during inflation can be ensured if one includes
a (positive) non-minimal coupling $\xi$ between the Ricci scalar and
the Higgs field 
$\L \supset \frac{1}2 \mpl^2 R (1-2\xi \Phi^\dagger \Phi)$
\cite{rajantie1,higgstory,Kamada:2014ufa}, as this induces an
effective stabilizing Higgs mass.  The vacuum is stable during
inflation if $\xi \gtrsim 3/8$, but even smaller couplings
$\xi \gtrsim 0.01$ are admissible depending on post-inflationary
evolution \cite{higgstory}. The non-minimal coupling to gravity is
allowed by all symmetries, in fact such a term will be generated by
loop effects, although it is not clear whether it will have the right
sign and size.

This is not the end of the story though. After inflation, as the
inflaton starts oscillating at the bottom of its potential, the
effective Higgs mass induced by the non-minimal coupling oscillates
between positive and negative values, which can give rise to efficient
production of Higgs quanta in a non-perturbative process called  preheating
\cite{Kofman1,Kofman2,Shtanov:1994ce,Felder:2000hj,Dufaux,Tsujikawa:1999jh,Bassett:1997az}.  The
produced Higgs quanta contribute to the effective Higgs mass and the
energy momentum tensor --- this is similar to temperature corrections
in a thermal bath, although the Higgs spectrum is highly non-thermal
--- and both effects can affect vacuum stability \cite{rajantie2}.
Preheating is efficient if
\be
q_0 \equiv \frac34\(\xi -\frac16\) \frac{\chi_0^2}{\mpl^2} \gtrsim \O(5),
\label{efficient}
\ee
with $\chi_0$ the inflaton field value at the end of inflation. In
this case particle production is explosive, and within a few inflaton
oscillations the produced quanta will completely dominate the Higgs
mass. As this all happens at high scales $ H > \mu_{\rm cr}$ the quantum
contribution to the effective mass is negative and destabilizes the
vacuum \cite{rajantie2,ema,Kohri:2016wof,Kohri:2016qqv,Enqvist:2016mqj}.

For smaller $q_0$-values results are less clear, and there is no
consensus in the literature on the fate of our vacuum, mainly because
different criteria for stability are used
\cite{rajantie2,ema,Kohri:2016wof,Kohri:2016qqv,Enqvist:2016mqj}. Even
though in this case particle production is not efficient enough
initially to destabilize the vacuum, this may still happen eventually
as the classical potential red shifts away faster than the quantum
corrections due to Higgs production. If the Higgs mass is dominated by
the production term only after the Hubble scale has dropped below the
critical value $H < \mu_{\rm cr}$, the quantum contribution to the
effective mass is positive around the maximum of the potential, and
thus will only raise the barrier separating the true and false vacuum.
Nevertheless, tunnelling to the true vacuum may be enhanced by the
non-zero energy density.  Although a full calculation is needed to
assert this effect, we expect decay to be fast when the energy density
becomes of the order of the maximum of the potential.  For small $q_0$
we will use this as our criterion for stability.

Arguably the most interesting part of parameter space is
$q_0 \lesssim 1$. From a naturalness point of view order one couplings
$\xi$ are favored, and for both chaotic and Starobinksy type inflation
models $\chi_0 \sim 1$, which implies $q_0 = \O(1)$.  One of the main
points of this paper is that exactly for these $q_0$-values the
effective Higgs mass and especially the energy density will depend
sensitively on the choice of vacuum, and no definite statements about
stability can be made.

The Higgs mass, which gets a one-loop correction proportional to the
Higgs two-point Green function, as well as the energy density can be
split in a divergent vacuum piece, which is absorbed in counterterms
and leads to the running of the couplings, and a finite piece due to
Higgs excitations on top of the vacuum. Different choices of vacuum
can be viewed as different renormalization schemes, as they lead to
different renormalization conditions defining the physical
couplings. In a cosmological setting where the background energy
density in the inflaton field is changing with time, the vacuum choice
is not well-defined.  Usually, the zeroth order adiabatic vacuum is
chosen. In the asymptotic regions where the system is (nearly)
time-independent --- at initial times during slow roll inflation and
at final times after inflaton decay --- it reduces to the usual in/out
vacua.  Moreover, if the background is slowly changing with time the
production of high momentum modes is exponentially suppressed,
supporting the vacuum interpretation \cite{BirrellDavies}.

The high momentum modes are adiabatic during at least part of the
inflaton oscillation, and these moments in time can be used to
evaluate the Green function. The problem with the current set-up is
that smaller momentum modes $k<k_n$ violate the $n$th order
adiabaticity condition and are never adiabatic. Since $k_n$
increases with time, the contribution of these non-adiabatic modes to
the Higgs mass and energy density increases with time until they fully
dominate the result. A direct way to monitor the vacuum dependence is
to compare quantities calculated using the zeroth and second order
adiabatic vacuum.

Of course, the electroweak vacuum is either stable or unstable, this
cannot depend on the choice of vacuum.  However, the problem is in
defining the renormalized couplings in the theory --- which is
necessary to find a critical coupling below which the vacuum is stable
--- as different vacuum choices correspond to different counterterms.
As usual this freedom can be fixed by measuring the
couplings. However, this measurement has to be done {\it during}
preheating, since different adiabatic vacua only give different results
during preheating and not in the adiabatic initial and final states.
Moreover, even if such a measurement was in principle possible, one
would also need to use non-adiabatic methods to calculate the Green
function for the results to be useful.

This paper is organized as follows. In the next section we introduce
the Lagrangian and derive the equations of motion for the inflaton and
Higgs field.  In section \ref{s:quantum} we then discuss the one-loop
corrections to the effective potential and energy density, which are
defined via an adiabatic renormalization scheme.  Semi-analytic approximations of the effective Higgs mass and energy density are given in section \ref{s:analytic}, as well as a comparison with the numerical results. In section \ref{s:adiabatic} we study the adiabaticity conditions and compute the
vacuum dependence of the quantities of interest. In section \ref{s:vacstab} we formulate our criteria for vacuum stability, discuss relevant time scales and corroborate our analytic results with numerical calculations. We end with concluding remarks.

\paragraph{Notation.}
We use a metric that is mostly positive $(-,+,+,+)$, and we set the
reduced Planck mass to unity $\mpl^2 = (8\pi G_N)^{-1} =1$.

The relevant equations depend on the combination $(\xi -1/6)$, for
which we introduce the shorthand notation
\be
\hat \xi \equiv  \(\xi -\frac16\).
\label{hatxi}
\ee

Time is measured in number of inflaton oscillations $T$.  The
frequency of the Higgs perturbations is periodic with frequency
$\delta T =\frac12$. To indicate a particular time during an oscillation,
we use the notation $T \is \frac14$, meaning $T = \frac14$ mod $\frac12$.

The inflaton background is an oscillating function with a
time-dependent amplitude $A(T)$.  The initial conditions for the
inflaton field amplitude and scale factor at $T=0$ are denoted by 
$A(0)=A_0$ and $a(0)=a_0$. We will also need the amplitude and scale
factor at $T=1/4$, which are denoted by $A(\frac14) = A_{1/4}$ and
$a(\frac14)= a_{1/4}$. Finally, after a few oscillations the amplitude
and scale factor are well approximated by $A(T) = A_T/T$ and
$a(T) = a_T T^{2/3}$, with normalization constants $A_T$ and $a_T$.
The values for all of them are taken from the numerical solution of
the classical background.  For future reference we list them
here
\be
\{A_0,\, a_0\} =\{1/2,\, 1\},\quad  \{A_{1/4},\, a_{1/4}\} \approx \{0.1,\, 1.3\},\quad 
\{A_T,\, a_T \} \approx \{0.25,\, 1.7\}.
\label{Aval}
\ee
%


\section{Classical action}
\label{s:classical}

We study a Higgs field that is non-minimally coupled to the Ricci
scalar. We are interested in the behavior during preheating, the
period just after the end of inflation when the inflaton is
oscillating in its potential, and the inflaton field still dominates
the energy density.  The Lagrangian is given by\footnote{ In principle
  one should also add a quartic interaction term between the inflaton
  field $\chi$ and the Higgs field
  $\L\supset \sqrt{-g}\, \kappa^2 \chi^2 |\Phi|^2$, which is allowed
  by the symmetries.  We will assume that the coupling $\kappa$ is
  small, and neglect this term.}
\be
\frac{\L}{\sqrt{-g}} =  \frac{1}{2} \(1 -  2\xi \Phi^\dagger
  \Phi\) R +\L_{\rm SM} + \L_{\rm inf},
\ee
with $\Phi$ the SM Higgs doublet.

\subsection{Inflaton background}
For the inflaton we take a quadratic potential 
\be
\frac{\L_{\rm inf}}{\sqrt{-g}} = -\frac12 (\partial \chi)^2 - \frac12
m_\chi^2 \chi^2.
\ee
This is a good approximation for many inflationary models soon after
the end of inflation, as for small field values $\chi \ll 1$ the
potential is generically dominated by the quadratic term.  However,
since most of the Higgs fluctuations are produced during the first few
inflaton oscillations, where deviations from the quadratic potential
can be significant, for example in Starobinksy inflation, more precise
calculations may require a model dependent inflaton potential.

The inflaton dominates the energy density in the universe.  The Hubble
constant and Ricci scalar can then be expressed in terms of the
inflaton field:
\be
3H^2 \simeq \frac12 \dot \chi^2 + \frac12 m_\chi^2 \chi^2,
\qquad
R = 6 \( \dot H + 2 H^2\) 
\simeq -\dot \chi^2 +2 m_\chi^2 \chi^2.
\ee
The equation of motion for the inflaton is:
\be
\ddot \chi + 3 H \dot \chi + m_\chi^2 \chi =0.\label{eominf}
\ee
We choose initial conditions at time $t_0 =2\pi/m_\chi$ for the
inflaton field and scale factor
\be
A_0 \equiv \chi(t_0) = \frac{1}{2}, \quad \dot \chi(t_0) =0, \quad
a(t_0) =1.
\label{initial_infl}
\ee
We take $m_\chi = 10^{-5}$ for the inflaton mass, which is the right
order of magnitude for both chaotic inflation with a quadratic
potential and Starobinsky inflation; in both models $\chi \sim 1$ at
the end of inflation, in agreement with \eref{initial_infl}.

After a few oscillations the inflaton background is very well
approximated by a periodic cosine function with a decreasing
amplitude\footnote{The phase is actually shifted at late times and a
  better approximation is
  $\chi \approx (A_T/T) \cos(2\pi T-\frac{\pi}{8} )$. However, the
  phase is irrelevant for most considerations, and for simplicity we
  drop it.}
\be
\chi \approx  \frac{A_T \cos(2\pi T) }{T},
\label{chiT}
\ee
with amplitude 
$A_T \approx 0.25$, which is determined by a fit to the numerical solution.
The number of inflaton oscillations $T$ is approximated by
\be
T \approx  \frac{m_\chi t}{(2\pi)}-1.
\label{T}
\ee
After
the first few oscillations the inflaton starts to behave as a cold
dark matter fluid, i.e., averaged over an oscillation the inflaton
has zero pressure and the energy density red shifts as
$\rho_\chi \propto a^{-3}$.  The scale factor grows as
$a \propto t^{2/3}$, which can be written as
\be
a \simeq a_T T^{2/3},
\label{aT}
\ee
with $a_T \approx 1.7$.  At late times the Ricci scalar and
Hubble constant evolve as
\be
R = \frac{A_T^2 m_\chi^2 (1+3\cos(4\pi T))}{2 T^2} + \O(T^{-3}),
\qquad
H^2 = \frac{A_T^2 m_\chi^2}{6 T^2}+ \O(T^{-3}).
\label{RH}
\ee

\subsection{Mode equation for the Higgs field}

The relevant part of the Lagrangian for the production of Higgs modes is
\be
\frac{\L}{\sqrt{-g}} =  \frac{1}{2} \(1 -  2\xi |\Phi|^2
  \) R - |D_\mu \Phi|^2 -V +...
\ee
with the Higgs potential
\be
V =  -\mu^2 |\Phi|^2 + \lambda |\Phi|^4.
\label{V}
\ee
During preheating the Higgs mass $\mu^2$ is very small compared to
the effective mass generated by the coupling to $R$, and will be
neglected in our computations.  We concentrate on the production of
the radial Higgs field, and neglect all other SM particles. In unitary
gauge $\Phi^\dagger \Phi=\Phi_R^2/2$, with $\Phi_R$ a real scalar that
can be split in a background field plus fluctuations:
\be
\Phi_R(\vec x,t) = \phi(t) + \vp(\vec x,t).
\ee
We are interested in the regime of small Higgs field values
$ \xi \phi^2 \ll 1$, then to leading order the results are the same in
the Einstein and Jordan frame.  This allows to treat gravity as a
classical background.  The FRW metric for a homogeneous isotropic
universe is
\be
\dd s^2 =-\dd t^2 + a^2(t) \dd \vec x^2 = a^2(\tau) \(-\dd \tau^2 +\dd
\vec x^2 \).
\ee
The conformal time $\tau$ is defined via $\dd\tau = \dd t/a(t)$.
Derivatives with respect to coordinate time $t$ are denoted by an overdot
and derivatives with respect to conformal time by a prime.

Using conformal time and defining `conformal'  fields 
\be
\bar \phi = a \phi, \quad
\bar \vp = a \vp, 
\quad \bar V = a^4 V,
\ee
the action for the fluctuations becomes 
(where we neglected the quartic
self-interaction term)
\begin{align}
S &\supset \frac12 \int \dd^3x \dd t a^3 \[ \dot \vp^2 +\frac{(\del \vp)^2}{a^2}
- \xi R \vp^2- V_{\phi\phi} \vp^2 \]
\nn \\
&=\frac12 \int \dd^3 x \dd \tau \[ \(\partial_\tau \bar \vp\)^2
  + (\del \bar \vp)^2 - M^2 \bar \vp^2
 - 2\frac{a'}{a} \bar \vp \partial_\tau \bar \vp + \(\frac{a'}{a}\)^2 \bar \vp^2 
 -\frac1{6}
  a^2 R \bar \vp^2\] \nn \\
&=\frac12 \int \dd^3 x \dd \tau \[ \(\partial_\tau \bar \vp\)^2
 + (\del \bar \vp)^2 - M^2 \bar \vp^2 \],
\label{action}
\end{align}
with the background dependent, and thus time-dependent, effective mass
term
\be
  M^2 = a^2\( \(\xi -\frac{1}{6}\)R(t) + V_{\phi\phi}\) .
\label{M}
\ee
In the second step we switched to conformal time and fields.  To get the
final expression we integrated by parts and further used that
$(a'/a) = aH$, $R= 6(\dot H + 2H^2)$, and
$\partial_\tau (a'/a) = (aH)^2+a^2 \dot H$.  The system is now
equivalent to that of a harmonic oscillator with a time dependent
frequency in Minkowski space, and we can apply the usual flat space
methods.  The field $\bar{\vp}$ can be expanded in mode functions
\be
\bar \vp = \int \frac{\dd^3 \vec k}{(2\pi)^3}
\( a_{\vec k} U_k(t) \e^{i \vec k.\vec x} 
+a^\dagger_{\vec k} U^*_k(t) \e^{-i \vec k.\vec x} 
\),
\ee
that satisfy the mode equation:
\be
U_k'' + \omega_k^2U_k = 0, \quad \omega_k^2 = k^2 + M^2.
\label{mode}
\ee
%

The term
$a^2 V_{\phi\phi}= \bar V_{\bar \phi \bar \phi} = 3\lambda \bar\phi^2$
in the effective mass \eref{M} will be neglected, because it only
becomes important once the classical Higgs field is close to the value
at the maximum. For smaller field values, it does not play an
important role. In order to find out \emph{whether} the classical
field obtains values as big as the value at the maximum, we thus do
not need the effect of the $3\lambda \bar\phi^2$-term. Our
approximation becomes unreliable for larger field values, but that is
not the regime that we are interested in.

\section{Quantum effective action}
\label{s:quantum}

In the previous section we outlined the behavior of the classical
inflaton field, which dominates the energy density, and gave the
classical mode equations for the Higgs fluctuations. In this section
we will discuss the one-loop corrections to the effective action and
mode equation for the Higgs field, and the contribution of the Higgs
modes to the energy density. 

In non-equilibrium systems we are mostly interested in expectation
values, which can be calculated using the CTP formalism
\cite{Keldysh:1964ud,Schwinger:1960qe,Jordan:1986ug,Calzetta:1986ey,Calzetta:1986cq}.
The effective action and energy density only depend on the equal-time
Green function, which we define with appropriate boundary
conditions. More details can be found for example in \cite
{Ringwald:1986wf,Ringwald:1987ui,Baacke:1997rs,Baacke:1996se}.

\subsection{Green Function}

We start off by defining the Higgs Green function, which will enter the
quantum corrected effective potential. As follows from the action for
the fluctuations \eref{action}, the rescaled Green function
$\bar G(\tau,x;\tau',x') \equiv \langle T \bar \vp(x,\tau) \bar
\vp(x',\tau') \rangle$ satisfies the Green function equation
\be
\[\partial_\tau^2 -\del^2 + M^2\] \bar G(\tau,x;\tau',x')=-i
\delta(\tau,x;\tau',x').
\label{G1}
\ee
Now we Fourier transform $\bar  G(\tau,x;\tau',x') = \int \frac{\dd^3
  k}{(2\pi)^3}
 \bar G_k(\tau,\tau')
\e^{i\vec k. (\vec x-\vec x')}$, and make the Ansatz
\be
\bar G_k(\tau,\tau') =  c \[ U_k(\tau) U_k^*(\tau')\Theta(\tau -\tau') +
U_k^*(\tau) U_k(\tau')\Theta(\tau' -\tau)\],
\ee
with $\Theta(\tau -\tau')$ the Heaviside step function.
Substituting this into \eref{G1}, and using the mode equation
\eref{mode}, we find
\be
c = -\frac{i}{\W_k}, \quad {\rm with} \quad \W_k = U_k'(\tau) U_k^*(\tau) - U_k^{*'}(\tau)U_k(\tau).
\ee
The Wronskian is time-independent, $\partial_\tau \W_k =0$, and is fixed
by the initial conditions:
\be
U_k(0) = u_k,\quad U_k'(0) = - i \wi U_k(0), \quad 
\W_k = -2i \wi |u_k|^2,
\ee
with $\wi = \omega_k(\tau_0)$.  We will take $u_k =1$. Note that
different normalizations are used in the literature (e.g.
$u_k = 1/\sqrt{2\wi}$
\cite{rajantie2,ema,Baacke:1997rs}); however, in all
cases the ratio $U_k(\tau_0)/U'_k(\tau_0)$ is the same, which assures
that the normalization drops out of the final result, and is thus
arbitrary.  The effective potential only depends on the equal-time Green function. In terms of the mode functions \cite{Baacke:1997rs,Baacke:1996se}:
\be
\bar G(\tau) = \langle \bar\varphi(\tau)^2 \rangle= -\int \frac{\dd^3
  k}{(2\pi)^3} \frac{i}{\W_k} |U_k|^2
= \int  \frac{\dd^3
  k}{(2\pi)^3} \frac{1}{2\wi} |U_k|^2.
\label{G2}
\ee
In the rest of this paper we are only interested in the equal-time
Green function, and for notational convenience the explicit
time-dependence of $\bar G$ is often dropped.

\subsection{Energy density}

The fluctuations $U_k$ give a contribution to the energy-momentum
tensor.  The energy density can be split into a classical part
(dominated by the inflaton contribution) and a quantum part.  We are
interested in the energy density of the Higgs fluctuations in the
Einstein frame.  There are two equivalent ways to derive this: either
work in the Jordan frame and treat the Ricci scalar as a classical
background source, or perform a conformal transformation to the
Einstein frame and derive the energy density in that frame.  We use the former method, but we stress that both methods give exactly the
same result in the small field limit $\xi \phi^2 \ll 1$.  The action
for the Higgs fluctuations was given in \eref{action}.  In terms of
the mode functions the (conformal) energy density derived from this
action is
\begin{align}
  \bar{\rho} &= \int \frac{d^3 k}{(2\pi)^3} \bar \rho_k= \int \frac{d^3 k}{(2\pi)^3} \frac{1}{4\omega_{k,0}} 
\left\{ |U'_k|^2 + \omega_k^2 |U_k|^2  \right\}.
\end{align}
%

\subsection{Adiabatic Renormalization of $\bar G$ and $\bar\rho$}
\label{AdRen}

The (equal-time) Green function $\bar G$ and the energy density
$\bar\rho$ are UV-divergent and need to be regularized. A convenient
method for renormalization in an expanding universe is the method of
adiabatic renormalization
\cite{BirrellDavies,Parker:1974qw,Fulling:1974pu,Bunch:1980vc}.  The
renormalized quantities are defined by subtracting the $n$th order
adiabatic approximation of the quantity of interest from the divergent
expression. The quantity of interest is thus defined with respect to
the $n$th order adiabatic vacuum.  This renormalization procedure is
particularly easy to implement, and widely used in preheating studies,
and thus also in the Higgs studies
\cite{rajantie2,ema,Kohri:2016wof,Kohri:2016qqv,Enqvist:2016mqj}. The
downside of this method is that the renormalization conditions for the
couplings and fields in the theory are only implicitly defined, making
it harder to define the renormalized couplings \cite{tranberg}.  In
fact, although for the non-interacting scalar theory adiabatic
subtraction is equivalent to redefining the constants of the original
action \cite{Bunch:1980vc}, new counterterms are needed in the
interacting non-equilibrium theory \cite{tranberg,Paz:1988mt}.

The Higgs mode functions behave adiabatically if the frequency
satisfies the adiabaticity conditions
\be
\eps_n \equiv \left|\frac{\partial_\tau^n\omega_k}{\omega_k^{n+1}}\right| \ll 1 .
\label{adiabaticity}
\ee
Usually only the first two conditions $n=1,2$ are considered, but 
we will show that it is important to look at the full tower.  In the
adiabatic limit, the solution to the mode equation \eref{mode} can be
approximated by the WKB-solution:
\be
  v_k = \sqrt{\frac{\omega_{k,0}}{W_k}}\, \e^{-i\int^\tau
    W_k(\tau')d\tau'},
\label{vk}
\ee
where $W_k$ satisfies the non-linear equation
\be 
W_k^2(\tau) = \omega_k^2(\tau) 
-\frac12\( \frac{W''_k}{W_k} -\frac32 \frac{{W'_k}^2}{W_k^2}\).
\ee
In a slowly varying spacetime, this can be solved iteratively.
The zeroth and second order WKB-solutions are given by:
\begin{align}
  \left(W_k^{(0)}\right)^2 &= \omega_k ^2, \nn \\
  \left(W_k^{(2)}\right)^2 &= \omega_k^2 -\frac{1}{2}
                             \left(\frac{\omega''_k}{\omega_k}-\frac{3}{2}
                             \frac{\omega_k'^2}{\omega_k^2} \right).
\label{secondorder}
\end{align}
In general, the difference between the $n$th and $(n\!+\! 2)$th
frequency $W_k^{(n)}$ is of the order of the adiabaticity parameters
$\eps_{n+1}$ and $\eps_{n+2}$.  Expanding the Higgs field in the
adiabatic mode functions
\be
\bar \vp = \int \frac{\dd^3 \vec k}{(2\pi)^3}
\( a_{{\vec k}, {\rm ad}}^{(n)} v^ {(n)}_k(t) \e^{i \vec k.\vec x} 
+ a_{{\vec k}, {\rm ad}}^{(n)\, \dagger} v^{(n)*}_k(t) \e^{-i \vec k.\vec x} 
\)
\ee 
defines the adiabatic vacuum via
$ a_{{\vec k}, {\rm ad}}^{(n)} |0^{(n)} \rangle =0$
\cite{BirrellDavies}. All orders of the adiabatic vacuum reduce to the
usual in/out vacua in the static asymptotic regions. Since the high
momentum modes behave adiabatically, as the $k^2$-term is the dominant
term in $\omega_k$, the production of high momentum modes is
suppressed, as is expected in the vacuum.

Since large momentum modes are increasingly adiabatic, the exact mode
functions approach the WKB solution in the UV limit. It follows that
the divergences in $\bar G$ and $\bar{\rho}$ can be cancelled by
subtracting from the Green function and the energy density the
corresponding expression in the WKB-approximation:
\be
  \bar{G}^{(n)}_{\rm ren}= \bar{G} - \bar{G}^{(n)}_{\rm ad},
\qquad   \bar{\rho}^{(n)}_{\rm ren} = \bar{\rho} - \bar{\rho}^{(n)}_{\rm ad},
\ee
which gives
\be
  \bar{G}^{(n)}_{\rm ren} = \int \frac{d^3k}{(2\pi)^3}
  \frac{1}{2\omega_{k,0}} \left( |U_k|^2 - |v^{(n)}_k|^2\right) = \int
  \frac{d^3k}{(2\pi)^3} \left(\frac{1}{2\omega_{k,0}} |U_k|^2 -
    \frac{1}{2W^{(n)}_k}\right),
\label{Gren}
\ee
and
\be
 \bar{\rho}^{(n)}_{\rm ren}  =  \int \frac{d^3 k}{(2\pi)^3} \biggl [
  \frac{1}{4\omega_{k,0}} \left\{ |U'_k|^2 +
    \omega_k^2 |U_k|^2 
  \right\} 
-\frac{1}{4}\biggl\{ \left(\frac{1}{4}
  \frac{(W_k^{(n)'})^{2}}{(W^{(n)}_k)^3} + W^{(n)}_k \right) 
+\frac{\omega_k^2}{W^{(n)}_k}
\biggr\} \biggr].
\label{rho}
\ee
For the Green function the adiabatic subtraction term
$\bar G^{(n)}_{\rm ad}$ removes all divergences for $n \geq 0$.  Since
the degree of divergence is higher for the energy density it seems
that one has to go to higher order $n \geq 2$ to also remove the
log-divergence. However, since $\bar\rho^{(2m+2)} -\bar\rho^{(0)}$ is
finite for $m\geq 0$, the UV-behavior is the same for all orders;
this implies that the zeroth order adiabatic vacuum works just as
well for regularizing the integral.

In a time-dependent background the adiabatic vacuum $|0^{(n)} \rangle$
is not an eigenstate of the Hamiltonian, and it does not minimize the
energy density in adiabatic mode $v_k$ at a given time $\tau_p$ \cite{Winitzki}:
\be
\bar\rho^{(n)}_{k,\,{\rm ad}} (\tau_p)= \frac{1}{4} \left(\frac{1}{4}
  \frac{(W_k^{(n)'})^{2}}{(W^{(n)}_k)^3} + W^{(n)}_k 
+ \frac{\omega_k^2}{W^{(n)}_k}\right) 
= \frac12 \omega_k (\tau_p)+ \O\(\eps_{1},...,\eps_{m}\) 
\ee
Only when the adiabaticity conditions $\eps_n \ll 1$ are satisfied,
the adiabatic energy density is slightly higher than the minimum value
$\rho_k(\tau_p)|_{\rm min} = \frac12 \omega_k(\tau_p)$.
Thus for the k-modes that violate the adiabaticity condition the
adiabatic vacuum is not a good vacuum.

In many previous works on preheating in the Higgs system the adiabatic
particle number density was used as a measure for the efficiency of
preheating.  The zeroth order adiabatic particle number density can be
defined as $ n_k^{(0)} = \rho_k^{(0)}/\omega_k$.  This approach thus
has the same issues with the choice of vacuum.

The WKB-approximation cannot be used for negative frequencies.  The
renormalized Green function and energy density are only defined at
times for which $W_k^2 > 0$ and the lower bound $t_-$ of the integral
in the WKB-solution at time $t$ should be chosen such that $W_k^2 >0$
between $t_-$ and $t$.  To formally extend the Green function and
energy density in the tachyonic regions one can take absolute values
of the various terms in $\bar G_{\rm ad},\,\bar \rho_{\rm ad}$; this
is also what is (implicitly) done in the definition of adiabatic
particle number used in previous
studies \cite{rajantie2,ema,Kohri:2016wof,Kohri:2016qqv,Enqvist:2016mqj}.

\subsection{Effective potential}

To find the one-loop effective potential we use the tadpole method
\cite
{Ringwald:1986wf,Ringwald:1987ui,Baacke:1997rs,Baacke:1996se}. The
equation of motion for the conformal background field can be found by
requiring the tadpole to vanish: $\langle \bar\phi \rangle =0$.  This
gives
\be
\bar \phi'' + a^2(\xi-\frac16)R \bar \phi + \bar V_{\bar \phi} +
\frac12 \partial_{\bar \phi}( \bar m_{\bar \phi}^2) \bar G(\tau) + ... =0,
\ee
where ${G}(\tau)$ is the equal-time Green function of the conformal
Higgs field. Since the Higgs is the only field that is produced
directly during preheating, we neglect contributions to the equation
of motion of the other fields, as denoted by the
ellipses. Further, for the Higgs potential \eref{V} one has
$\bar V_{\bar \phi} \simeq \lambda \bar \phi^3$ and
$\partial_{\bar \phi}( \bar m_{\bar \vp^2}) = \bar V_{\bar \phi \bar
  \vp \bar \vp}$.
In a (nearly) static background, integrating the equation of motion
gives the usual Coleman-Weinberg correction to the effective
potential \cite{Coleman}.  Including explicit counterterms\footnote
{Since the non-minimal coupling is a non-renormalizable interaction
between gravity and the Higgs field, one cannot treat gravity as a
classical background in calculating the counterterms and RGE equation;
rather a covariant method should be used \cite{moss}.  Since the
details are not important for our purposes, we neglect this complication.}
 this gives
\be
 \bar V_{\rm eff} =  
\frac12 a^2 \(\xi + \delta \xi -\frac16\) R \bar \phi^2 +
\frac{(\lambda + \delta \lambda)\bar \phi^4}{4} + 3 \lambda \bar \phi^2 \bar
G(\tau) + ....
\label{CT}
\ee
Now we split the Green function in a vacuum part plus a part due to
excitations on top of the adiabatic vacuum
$\bar G = \bar G^{(n)}_{\rm ad} + \bar G^{(n)}_{\rm ren}$, according to the adiabatic
renormalization scheme \eref{Gren}. The divergent vacuum part can be
absorbed in the counterterms, together with the vacuum contribution of
all other SM particles.  In a static universe, defining appropriate
renormalization conditions this gives the standard results for the
renormalized couplings.  In an adiabatically expanding universe this
procedure can still be used at each moment in time since time derivatives
only give small corrections.  Although very easy to implement, the
disadvantage of the adiabatic renormalization scheme is that it is
hard to translate it into the explicit counterterms in the Lagrangian
\eref{CT} .  The effective potential can be renormalization group (RG)
improved to give
\be
 \bar V_{\rm eff} =  
\frac12 a^2 \(\xi(\mu)-\frac16\) R \bar \phi^2 +
\frac{\lambda(\mu)\bar \phi^4}{4} + 3 \lambda(\mu) \bar \phi^2 \bar
G^{\rm ren}(\tau).
\ee
with $\lambda(\mu),\, \xi(\mu)$ the running couplings.  In an
expanding universe the renormalization scale can be taken as the mass
of the top quark $m_t \sim \phi$, or if higher, the Hubble scale
\be
\mu \sim {\rm max}(H, \phi).
\label{mu}
\ee
The RGE for the quartic coupling and all other SM couplings are the
usual SM RGEs (see e.g.\cite{disc2}), the RGE for the
non-minimal coupling can be found in \cite{moss}.  The boundary
conditions are set by the measurement of the Higgs mass at the
electroweak scale, and the value of $\xi$ at the inflationary scale
that can be extracted from the CMB.  The term proportional to
$\bar G^{(n)}_{\rm ren}(\tau)$ is the contribution to the effective
potential due to Higgs quanta above the $n$th order adiabatic vacuum;
this is analogous to how the effective potential receives thermal
corrections in a plasma.

Different choices of vacuum, such as the zeroth and second order WKB
vacuum \eref{secondorder}, correspond to different counterterms, and
as a result the physical couplings are defined via different
renormalization conditions. Since the different order vacua all
coincide in the static asymptotic regions, the renormalization
conditions only differ during preheating.  As a result, to fully fix
the theory one has to measure the couplings during this period.  If
preheating is efficient, the vacuum dependence of the result is
relatively small and this is not a big problem.  However, as we will see,
if preheating is less efficient the vacuum dependence in the Green
function and energy density grows with time, i.e. the differences
$\bar G^{(0)} - \bar G^{(n)}$ and $\bar\rho^{(0)} - \bar\rho^{(n)}$ grow with
time.  Therefore, the ambiguity in the finite pieces of the
counterterms, and thus in the definition of the physical couplings via a
renormalization condition grows, and it becomes harder to extract a
reliable critical coupling for stability.  Note that the physics is
independent of our choice of vacuum, the electroweak vacuum is either
stable or not. The problem lies in the lack of a measurement of the
couplings during preheating, and secondly in the application of
equilibrium methods to a system that is not adiabatic.

We can read off the one-loop effective mass, which consists of the
tree-level term \eref{M} plus quantum correction due to the production
of Higgs quanta: 
\be
M_{\rm eff}^2 \simeq a^2 \(\xi(\mu)-\frac16\) R + 6 \lambda \bar
G^{\rm bg} (\tau)=  a^2 \hat \xi(\mu) R + 6 \lambda \bar
G^{\rm bg} (\tau)
\label{Meff}
\ee
where we neglected the subdominant tree-level contribution from the
quartic Higgs coupling, and used the notation \eref{hatxi}.

If the mass squared is positive at small field values --- either
because the tree-level mass dominates, or because the quartic coupling
is positive and the quantum correction to the mass is positive ---
the potential has a barrier separating the true and false vacua.
The maximum of the potential is
\be \bar V_{\rm max} = \frac14 \frac{M_{\rm
    eff}^4}{|\lambda|},
\label{Vmax}
\ee 
with $|\lambda| \sim 10^{-2}$.  Here we assumed
$H > \mu_{\rm cr} \sim 10^{11}$GeV (the critical scale where the
quartic coupling becomes zero $\lambda(\mu_{\rm cr}) =0$), in which
the coupling $\lambda(H)$ is negative and field independent near the
maximum.  For a smaller Hubble constant, there will be corrections as
the coupling $\lambda(\phi)$ is field dependent, but these corrections are small in the regime that we are interested in.

Along the same lines one can define the energy density of the Higgs
quanta on top of the adiabatic vacuum as $\rho^{(n)}_{\rm ren}$ as defined
in \eref{rho}, where the couplings can be taken as the running
couplings.

\section{Higgs effective mass and energy density}
\label{s:analytic}

In this section we will discuss the numerical results for the Higgs
effective mass and energy density, and develop a semi-analytical
understanding. More details on the numerical implementation are given
in section \ref{s:implementation}. The results are for the zeroth
order adiabatic vacuum. In the next section we look at the vacuum
dependence, and discuss the difference with the higher order vacua.

\begin{figure}
\begin{subfigure}{1\textwidth}
        \centering
      \includegraphics[width = 0.49\linewidth, bb = 0 4 359 225]{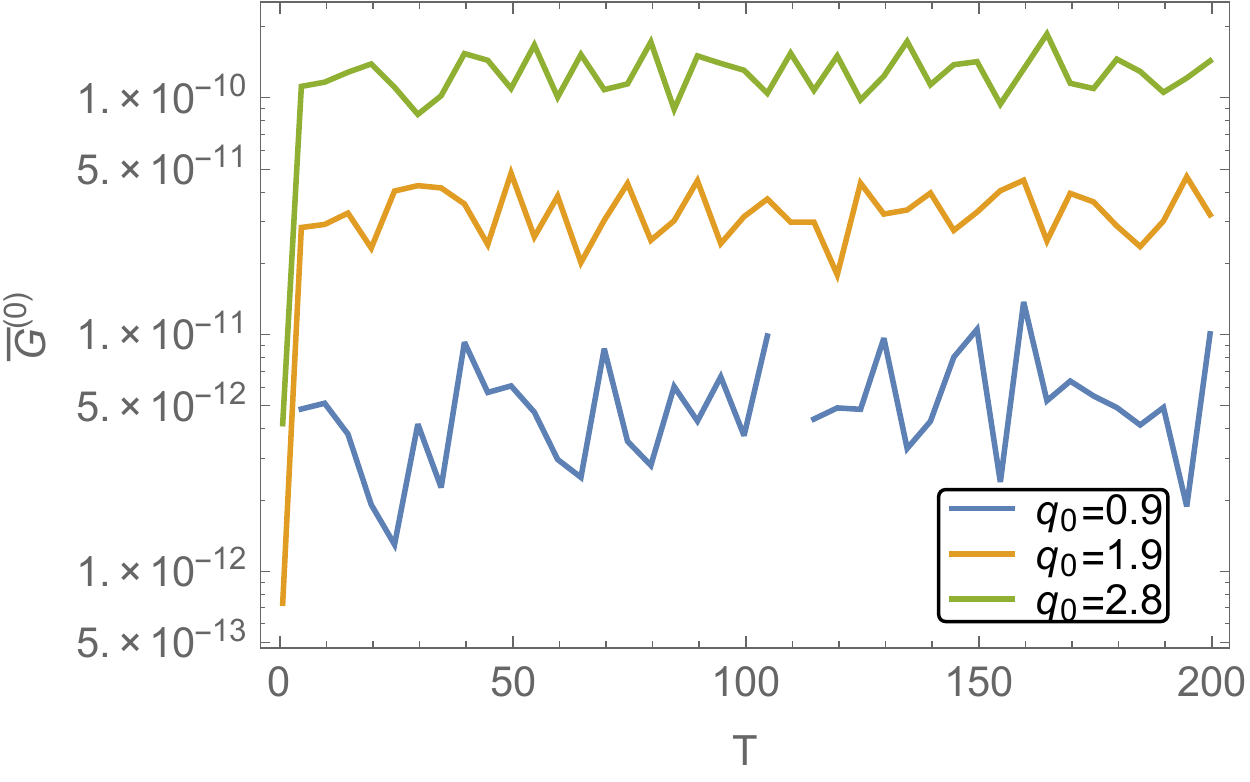}
 \includegraphics[width = 0.49\linewidth,bb = 1 3 359 222]{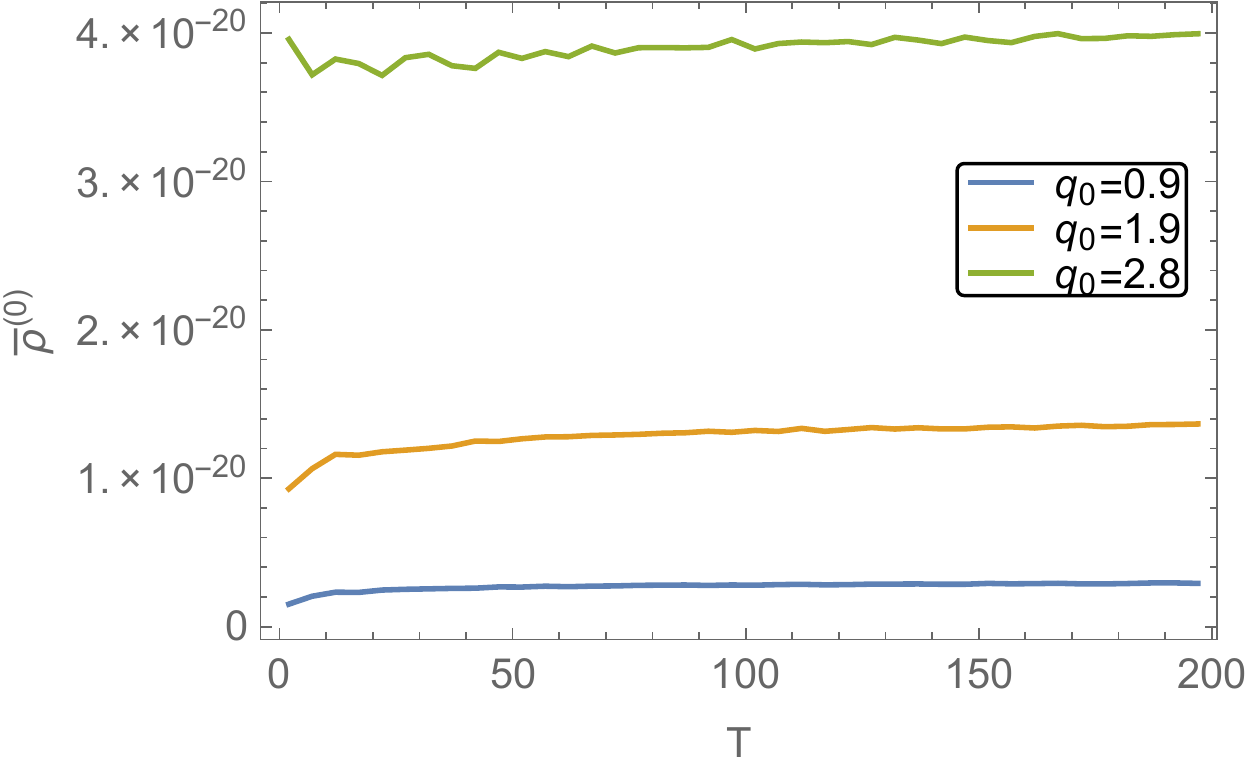}
        \caption{Left graph: $\bar G ^{(0)}$ as a function of time during the
  first 200 oscillations of the inflaton field. Right graph:
  $\bar \rho ^{0}$ during the first 100 oscillations. $\bar G ^{(0)}$
  and $\bar \rho ^{(0)}$ are plotted for $q_0 = 0.9$ (blue),
  $q_0 = 1.9$ (orange) and $q_0 = 2.8$ (green). The quantities are
  sampled with $\delta T = 5$, for values of $T$ where the vacuum dependence is minimal.}
\label{fig:largeT} 
    \end{subfigure} \\
    \begin{subfigure}{1\textwidth}
        \centering
\vspace{0.5cm}
 \includegraphics[width = 0.48\linewidth, bb = 1 4 358 239]{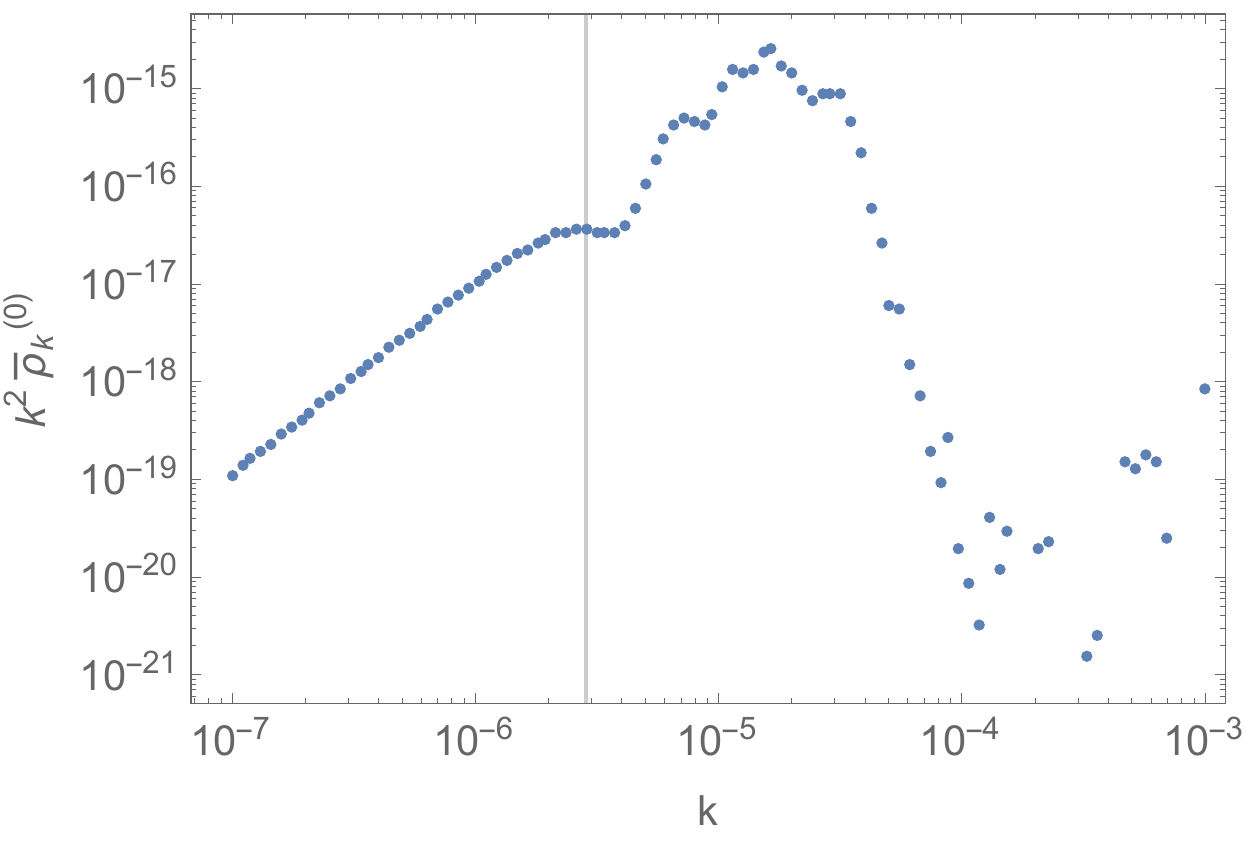}
\includegraphics[width = 0.48\linewidth,bb = 1 4 358 239]{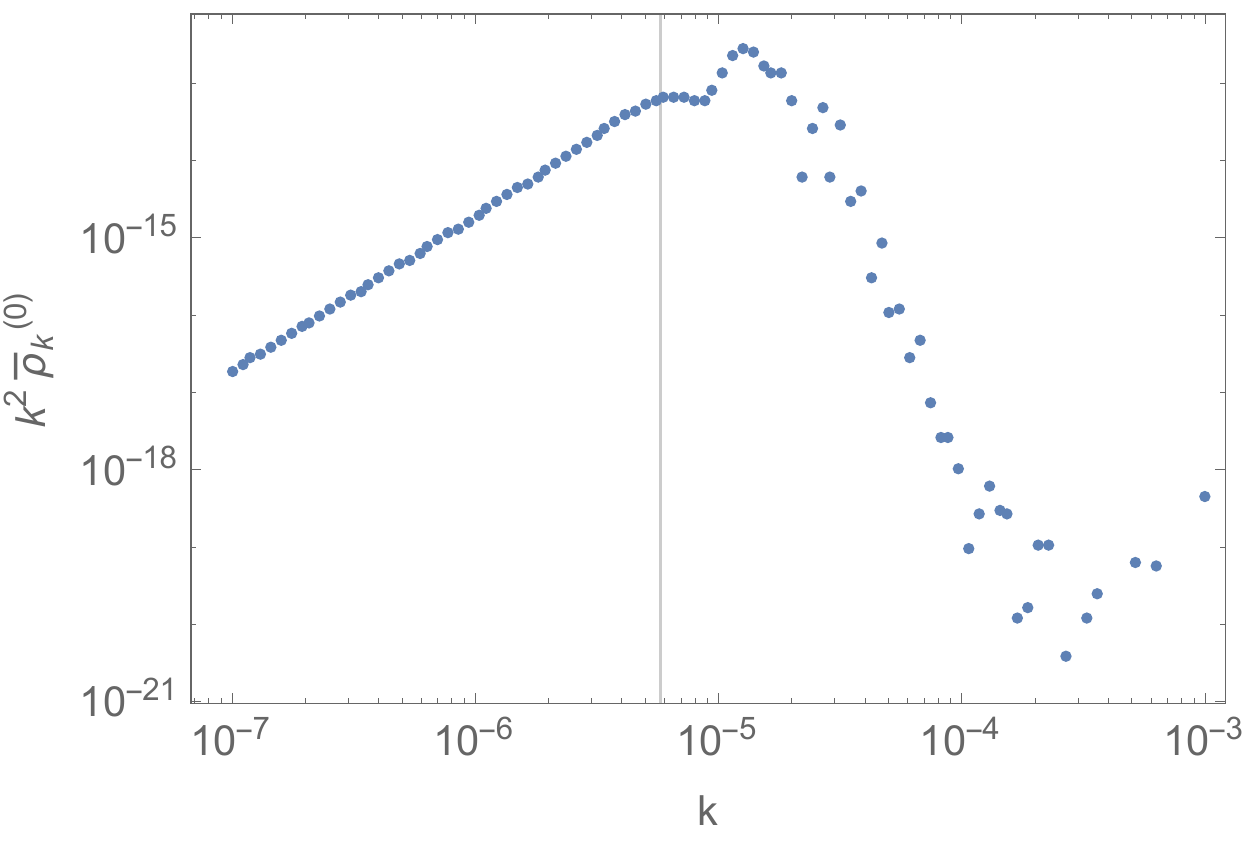}
        \caption{$k^2 \bar\rho_k^{(0)}$ as a function of $k$ at $T =
          3.5$ for $q_0 = 0.9$ (left graph) and $q_0 = 3.7$ (right
          graph). The vertical line indicates the value of
          $k_\text{tach}$.}
\label{fig:rho_k} 
    \end{subfigure}\\
    \begin{subfigure}{1\textwidth}
        \centering
\vspace{0.5cm}
      \includegraphics[width = 0.48\linewidth, bb= 0 1 356 239]{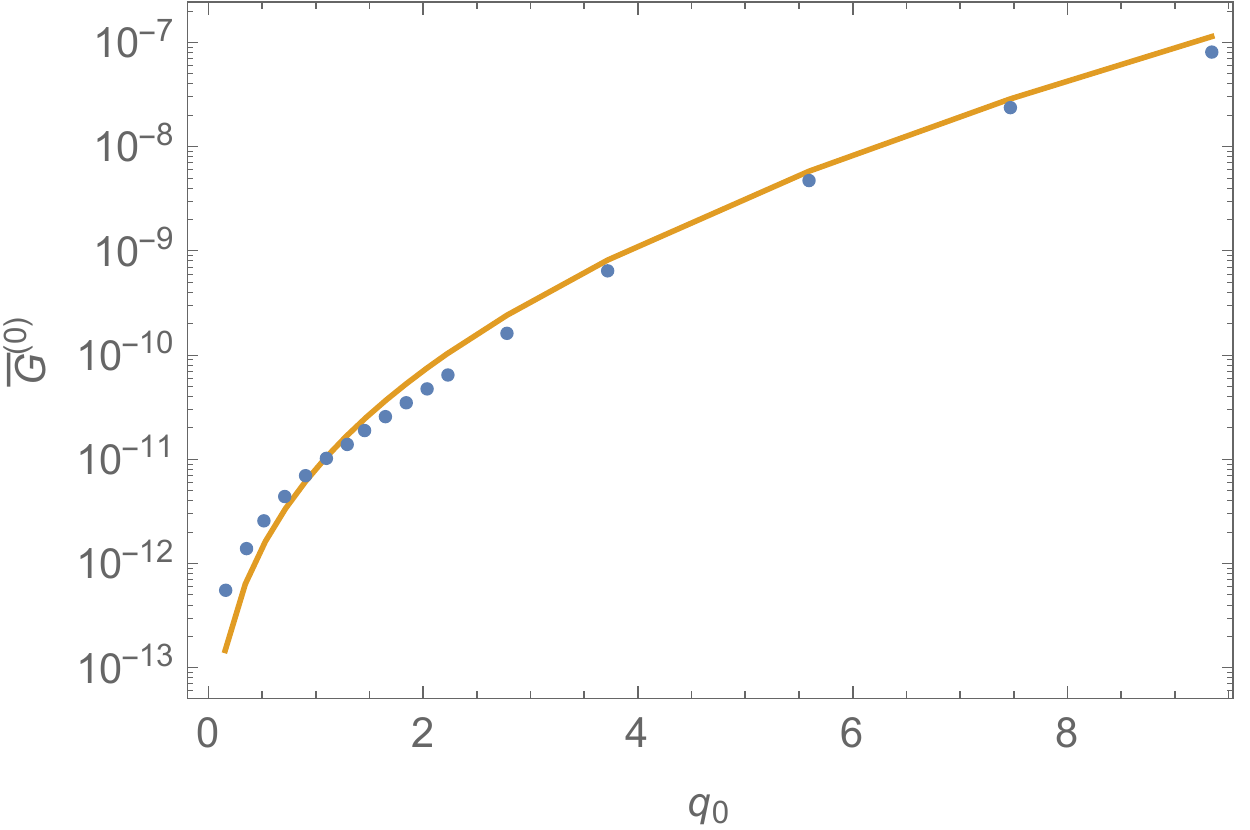}
  \includegraphics[width = 0.48\linewidth,bb= 1 1 356 239]{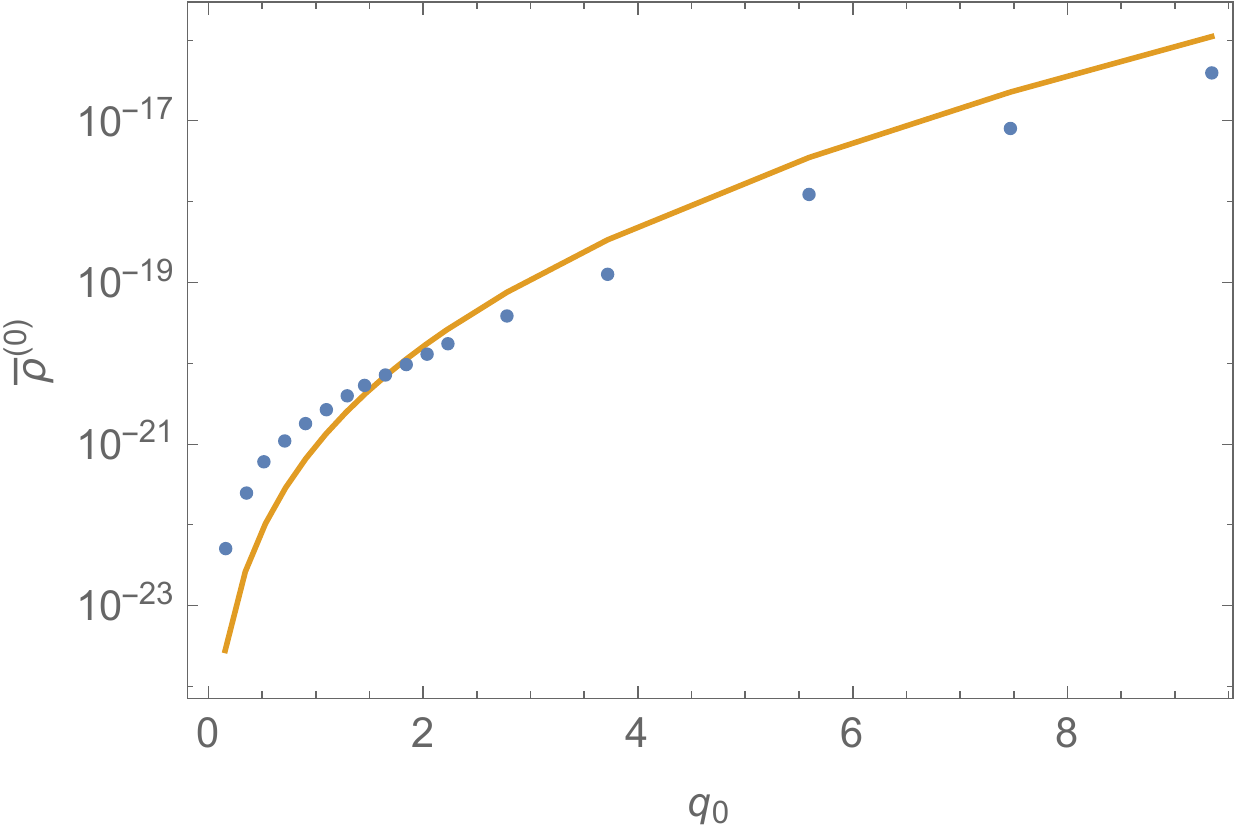}
        \caption{Left graph: comparison between the numerically computed values of $\bar{G}^{(0)}$ (blue) and estimate \eref{Gfit} (orange) as a function of
    $q_0$. $\bar{G}^{(0)}$ was computed after 1.5 oscillations, except
    for the two largest values of $q_0$, where it was computed after 3
    oscillations. Right graph: comparison between the numerically
    computed values of $\bar{\rho}^{(0)}$ (blue) and estimate
    \eref{rhofit} (orange) as a function of $q_0$. $\bar{\rho}^{(0)}$
    was computed after 3.5 oscillations.}
\label{fig:Grhofit} 
    \end{subfigure}
\caption{}
\end{figure}

\subsection{Green function}
 \label{s:Meff}

 For small enough comoving momentum $k$ the frequency squared
 $\omega_k^2$ of the Higgs modes oscillates between positive and
 negative values. Modes for which the frequency squared is negative
 are produced in a process called tachyonic preheating
 \cite{Felder:2000hj,Dufaux}.  Modes for which the frequency is
 non-adiabatic are produced in a process called parametric resonance
 \cite{Kofman1,Kofman2,Shtanov:1994ce}.  Most quanta are produced in
 the first few oscillations.

 We will estimate the contribution of the tachyonic and non-adiabatic
 modes to the Green function, guided by the numerical results.  The
 first step is to write the mode equation \eref{mode} in the form of
 the Mathieu equation. To do so we approximate the inflaton field and
 its derivative during the first oscillation as\footnote{After a few
   oscillations the amplitude is well approximated by
   $A(T) \approx A_T /T$ as in \eref{chiT}, but in the first few
   oscillations the amplitude decays faster and has a different time
   dependence.}
\be
\chi  \approx  A(T) \cos(2\pi T),
\quad
\partial_T \chi \approx A(T) \partial_T \cos(2\pi T ),
\label{smalltime}
\ee
that is, we neglect the time-dependence of the amplitude in taking the
derivative. If furthermore the time-derivative of the mode function is
approximated by
$\partial^2_\tau U_k \approx (a m_\chi/(2\pi))^2 \partial^2_T U_k$,
the mode equation \eref{mode} can be written in the form of a Mathieu
equation:
\be
\partial^2_z U_k + \(\A_k -2q \cos(2z)\) U_k =0,
\ee
with
\be
\A_k(T) = \frac{k^2}{a^2 m_\chi^2} + \frac12  \hat \xi A^2,
\quad
q(T) = \frac34 \hat \xi A^2,
\quad
2z= 4\pi T -\pi.
\label{q}
\ee
Note that the amplitude $A$, scale factor $a$, and thus the parameters
$\A_k,\,q$ are time-dependent in the expanding universe.  We define the
efficiency parameter as the $q$-value at the initial time
\be
q_0 \equiv  q(0) = \frac34  \hat \xi A_0^2.
\ee

Let's look at tachyonic preheating first. The small momentum modes are
tachyonic $\omega_k^2 < 0$ during part of the inflaton
oscillation. The frequency is most negative at times
$T \approx T_p =\frac14 + \frac12 p$, with $p$ integer.  It was
estimated in \cite{Dufaux} that during the $p$th burst of tachyonic
particle production, the tachyonic mode functions grow approximately
as:\footnote{The factor $({2\omega_{k,0}}/{\omega_{k}})$ enters
  because of our normalization of the mode functions.}
\be
|U_k(T_p)|^2 \sim \(\frac{2\omega_{k,0}}{\omega_{k}}\)
\e^{4x\sqrt{q_p}},
\label{Ugrow}
\ee
with $x$ an order one constant. Since $q$ decreases with time,
the production is dominated by the first few times the frequency becomes
tachyonic.
Consider then the first burst of particle production at $T_{p=0} =
1/4$.  Modes
with $k \lesssim k_{\rm tach}$ are tachyonic at that time, with
\be
k_{\rm tach}= 
\frac2{\sqrt{3}} a_{1/4} m_\chi \sqrt{q_{1/4}} ,
\label{ktach}
\ee
where we introduced the shorthand $q_{1/4} = q(T =1/4)$ and analogous
notation for other time-dependent quantities. The first burst of
production gives a contribution to the
Green function \eref{Gren}
\begin{align}
\bar G^{(0)} (1/4)& \sim \int^{k_{\rm tach}} \frac{\dd^3  k}{(2\pi)^3 2\omega_{k,0}}
|U_k(1/4)|^2 
\sim \frac1{2\pi^2} \int^{k_{\rm tach}} \dd k\,  k \e^{4x\sqrt{q_{1/4}}}
\sim \frac1{2\pi^2} k^2_{\rm tach}\e^{4x\sqrt{q_{1/4}}}. 
\end{align}
Here we used that the integral is dominated by the modes with
$k\sim k_{\rm tach}$, for which the approximation $\omega_k \sim k$ is
valid.  Particle production in the subsequent tachyonic intervals is
subdominant.  It follows that $\bar G^{(n)}$ approaches its asymptotic
value after about one inflaton oscillation, and remains constant after
that. The above approximation for the Green function is thus valid at
large times as well.  Relating $q_{1/4} = (A_{1/4}/A_0)^2 q_0 $, the
contribution of the tachyonic modes to the Green function can be approximated by:
\be
\bar G^{(0)} (T) 
\sim \frac1{2\pi^2} k^2_{\rm tach}\e^{4x\sqrt{q_{1/4}}} 
\sim 5 \times 10^{-13} q_0 \e^{c \sqrt{q_0}},
\label{Gfit}
\ee
where in the 2nd step we used $m_\chi = 10^{-5}$ and \eref{Aval}, and
defined $c =4x \sqrt{q_4/q_0}$.

We fit the growth factor $c$ to the numerical results
\be
c \approx 3.5.
\label{cval}
\ee
The left panel of figure \ref{fig:Grhofit} shows a comparison between
the numerically computed values of $\bar{G}^{(0)}$ and approximation
\eref{Gfit}. The approximation agrees with the numerical result within
a factor of two over the whole range $q_0 =0.15-9.3$, corresponding to
$\xi =1-50$.  The left panel of figure \ref{fig:largeT} shows the large time
behavior of the Green function in the zeroth order adiabatic vacuum.
The graphs show that $\bar G^{(0)} $ remains approximately constant
after the initial amplification, confirming our assumption that
production is completely dominated by the first few oscillations.

\subsection{Energy density}

Analogously to the Green function, the energy density is produced in the
first few oscillations and remains constant afterwards, as confirmed
by the numerical results shown in Figure \ref{fig:largeT}.

The contribution of the tachyonic modes to \eref{rho}  is estimated as 
\begin{align}
 \bar{\rho}^{(0)}(T) &\sim \frac1{8\pi^2} \int^{k_{\rm tach}} k d k \,
  \left\{ |U'_k|^2 +
\omega_k^2 |U_k|^2 
  \right\}  \Big|_{T =\frac14}
\sim 
 \frac1{4\pi^2} \int^{k_{\rm tach}} k^3 d k \,|U_k(1/4)|^2 
\nn \\
& \sim \frac{1}{4\pi^2}  k_{\rm tach}^4  \e^{c \sqrt{q_0}}.
\label{rhofit0}
\end{align}
In the 2nd
step we used that the modes $k \sim k_{\rm tach}$ dominate the
integral, analogously to the approximation made in the Green function.
Moreover, we assumed equipartition and used $|U'_k|^2 \sim k^2 |U_k|^2$,
which agrees with the numerical results. 

Numerically, we fit
\be
\bar{\rho}^{(0)} \sim 2.8 \times 10^{-23} q_0^2 \e^{c \sqrt{q_0}},
\label{rhofit}
\ee
which is about a factor $15$ larger than the contribution
from the tachyonic modes \eref{rhofit0}. The growth factor $c$ is given
by \eref{cval}.  This indicates that the energy density and Green
function are dominated by the same mode functions.

In the right panel of figure \ref{fig:Grhofit} a comparison between
the numerically computed values of $\bar \rho ^{(0)}$ after $3.5$
inflaton oscillations and approximation \eref{rhofit} is shown. For
$q_0 > 2$ the estimate and numerical result agree within a factor of
two, but for smaller $q_0$ estimate \eref{rhofit} underestimates the
numerical result by a larger margin.

There are several effects that can explain the factor 15 difference between the analytic and numerical result. First of all, the approximation $\omega_k \approx k$ could be slightly off. Secondly, the assumption that the production of modes after the first oscillation is negligible does not completely hold for larger $q_0$. Furthermore, we did not take into account the contribution of non-tachyonic modes.
Figure \ref{fig:rho_k} shows $k^2\bar\rho_k^{(0)}$ as a function of
$k$ at $T =3.5$. As expected, the resonance is most efficient for the
larger value of $q_0$. The right graph shows that for $q_0 = 3.7$ the
distribution is peaked at a $k_*$ that is slightly larger than
$k_\text{tach}$. The left graph shows that for $q_0 = 0.9$ the
difference between $k_\text{tach}$ and the most dominant mode $k_*$ is approximately a factor 5. This implies that there is a significant contribution of modes that are non-adiabatic, but never tachyonic. This leads to the conclusion that for $q_0 \lesssim 1$, parametric resonance as described in \cite{Kofman2} plays a role that is more important than tachyonic resonance. 

These three effects should in principle also apply to our estimate of $\bar G^{(0)} $. However, we did not need any multiplicative factor for our fit of $\bar G^{(0)}$, which suggests that the effects cancel in this case.

\section{Adiabaticity and vacuum dependence}
\label{s:adiabatic}

The adiabatic vacuum is a good vacuum for high-momentum modes, for
which the adiabaticity parameters defined in \eref{adiabaticity} are
small at least during part of the inflaton oscillation.  However, for
smaller momenta, adiabaticity is violated at all times. If these
momenta give a significant contribution to the Green function and/or
energy density, this introduces a large vacuum dependence, which
manifests itself in that different orders of adiabatic vacuum give
different results for $\bar G$ and $\bar \rho$.

The adiabatic vacua are all equivalent in the asymptotic time regions
where $\omega_k$ is (approximately) constant, but can differ
significantly during preheating.  Since different choices of vacua
correspond to different subtractions in $\bar G$ and $\bar \rho$, see
equations (\ref{Gren},~\ref{rho}), they correspond to different
counterterms and renormalization conditions.  There is thus an
uncertainty in the values of the renormalized couplings, which can
only be fixed (in theory) by a measurement during preheating.
Moreover, even if such a measurement was in principle possible, one
would also need to use non-adiabatic methods to calculate the Green
function for the results to be useful. Note that the physics is
independent of our choice of vacuum, the electroweak vacuum is either
stable or not. The problem lies in our theoretical description of this
process.

In this section we will discuss the adiabaticity condition and the
vacuum dependence of the Green function and energy density, which are
both input to determine the stability of the vacuum, as laid out in
section \ref{s:stability}.

\subsection{Adiabaticity conditions}

After a few oscillations the inflaton background is very well
approximated by expression \eref{chiT}.  The frequency for the
Higgs modes \eref{mode} becomes
\be
\omega_k^2 \simeq k^2 +\frac12 A_T^2 a_T^2 m_\chi^2 \hat \xi T^{-2/3} 
\(1+3\cos(4\pi T) \) + \O(T^{-5/3}).
\label{omegaT}
\ee
The frequency is periodic with $\delta T =1/2$.  In the rest of this
section we will use the notation $T \is 1/4$ to denote $T = 1/4$ mod
$1/2$.

Modes with $k < k_n$ violate the $n$th-order adiabaticity condition
$\eps_n > 1$, which involves $n$ derivatives of the
frequency.  Keeping only the leading term at large $T$ this gives
the estimate
\begin{equation}
\label{kn}
k_n \sim a_T m_\chi 
\( 4 \times 2^{(n-2)} \(\frac{A_T}{A_0}\)^2 q_0\)^{\frac{1}{(n+2)}} 
T^{\frac{2(n-1)}{3(n+2)}} \times
\left\{
\begin{array}{ll}
|\sin(4\pi T)|^{\frac1{(n+2)}},
&n \; {\rm odd}, \\
|\cos(4\pi T)|^{\frac1{(n+2)}},
&n \; {\rm even},
\end{array}
\right.
\end{equation}
with $A_T/A_0 \approx 0.2$, see \eref{Aval}.  Since even orders are
proportional to a cosine, and odd orders to a sine, $k_n$ cannot be
minimized simultaneously for all $n$.  At every moment during the
inflaton oscillations there are modes that violate some of the
adiabaticity conditions.  The momentum cutoff $k_n$ grows, and as time
goes by more modes become non-adiabatic.  At early times all $k_n$ are
similar for $q_0 = 1-10$, but the larger the order $n$ the faster
$k_n$ grows with time. Explicitly, parametrizing
$k_n\propto T^{\alpha}$ this gives $\alpha = \{0, 1/6,4/15,1/3\}$ for
$n= \{1,2,3,4\}$.

\begin{figure}
\centering
\begin{subfigure}{1\textwidth}
 \includegraphics[width = 0.48\linewidth, bb = 0 3 356 208]{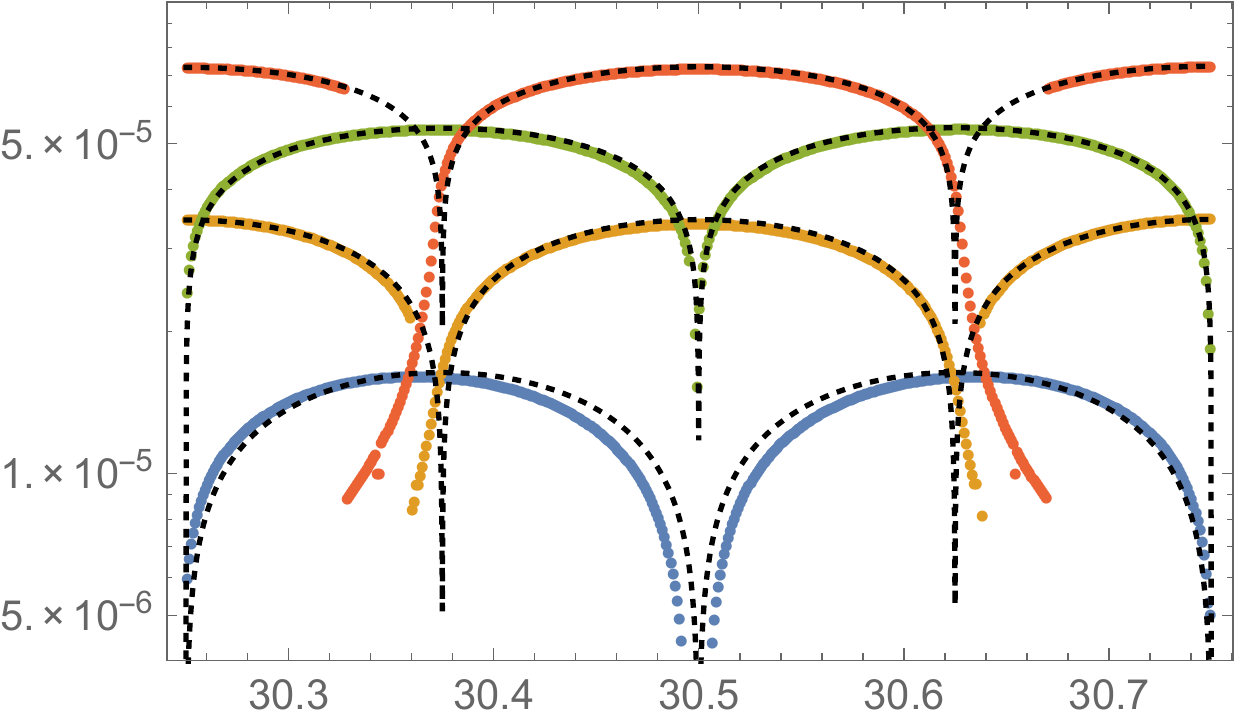}
  \includegraphics[width = 0.48\linewidth, bb = 0 3 356 205]{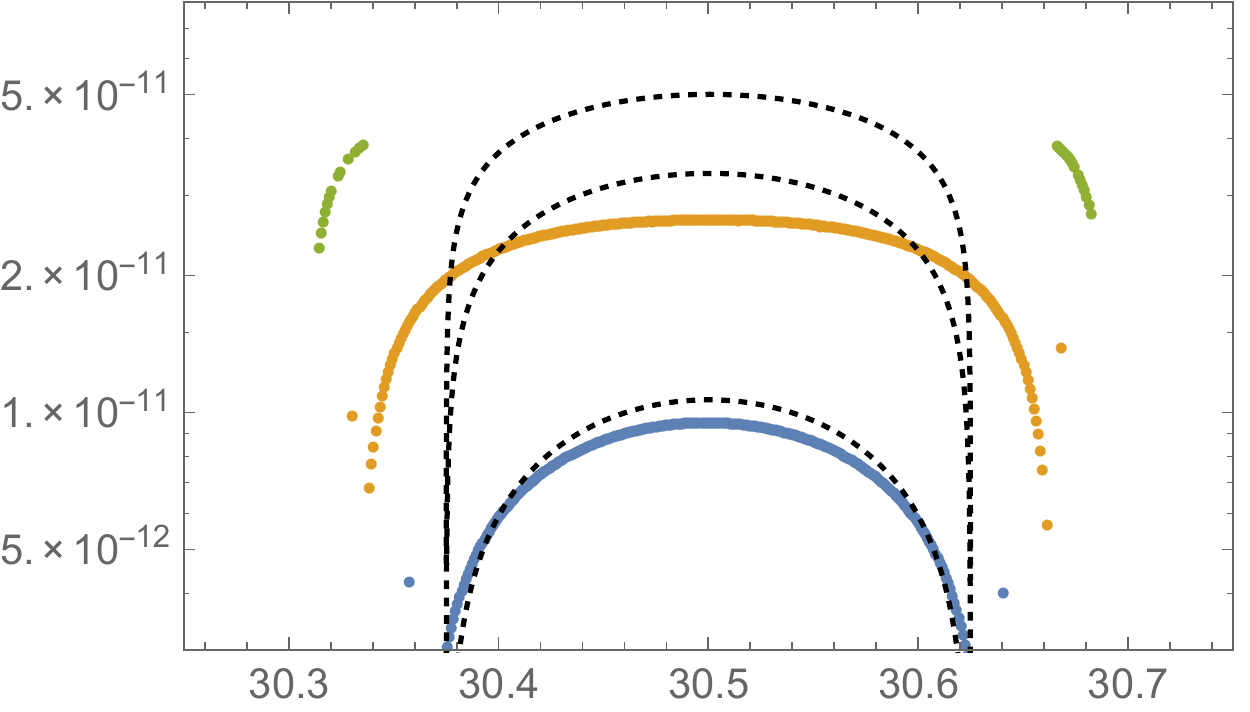}
\caption{Vacuum dependence.  The top left plot shows $k_n$ for
  which $\eps_n =1$ during a time interval $\delta T =1/2$. The blue,
  orange, green and red line corresponds to $n=1,2,3,4$ respectively,
  and the black dotted lines the analytical approximation
  \eref{kn}. The right plot shows $\Delta \bar G^{(n)}$; the blue,
  orange and green points correspond to $n=2,4,6$. Only points with
  ${\rm Im} (\Delta \bar G^{(n)}) < 0.01 {\rm Re} (\Delta \bar
  G^{(n)})$
  are shown.  The black dotted lines correspond to the approximation
  \eref{dG}. }
\label{F:vacuum1} 
\end{subfigure} \\
\begin{subfigure}{1\textwidth}
\vspace{1cm}
  \includegraphics[width = 0.48\linewidth, bb = 1 3 356 218]{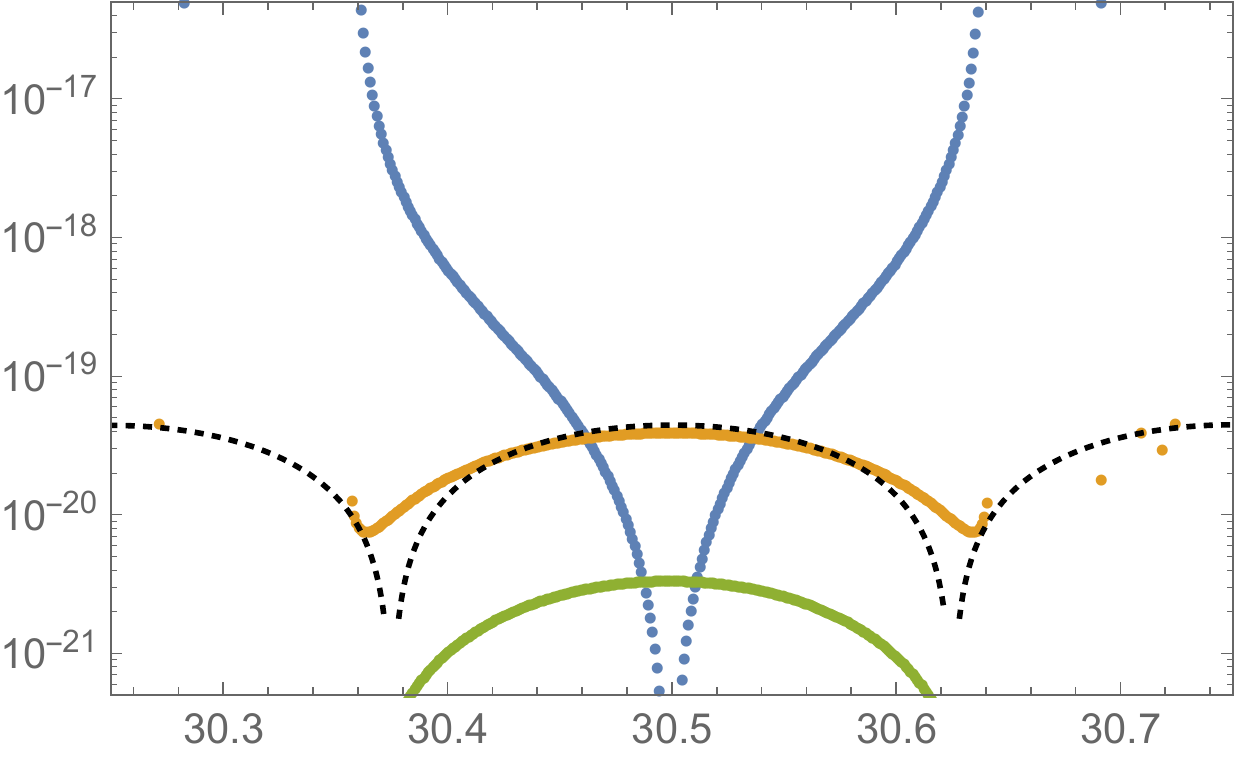}
  \includegraphics[width = 0.48\linewidth, bb = 1 3 356 218]{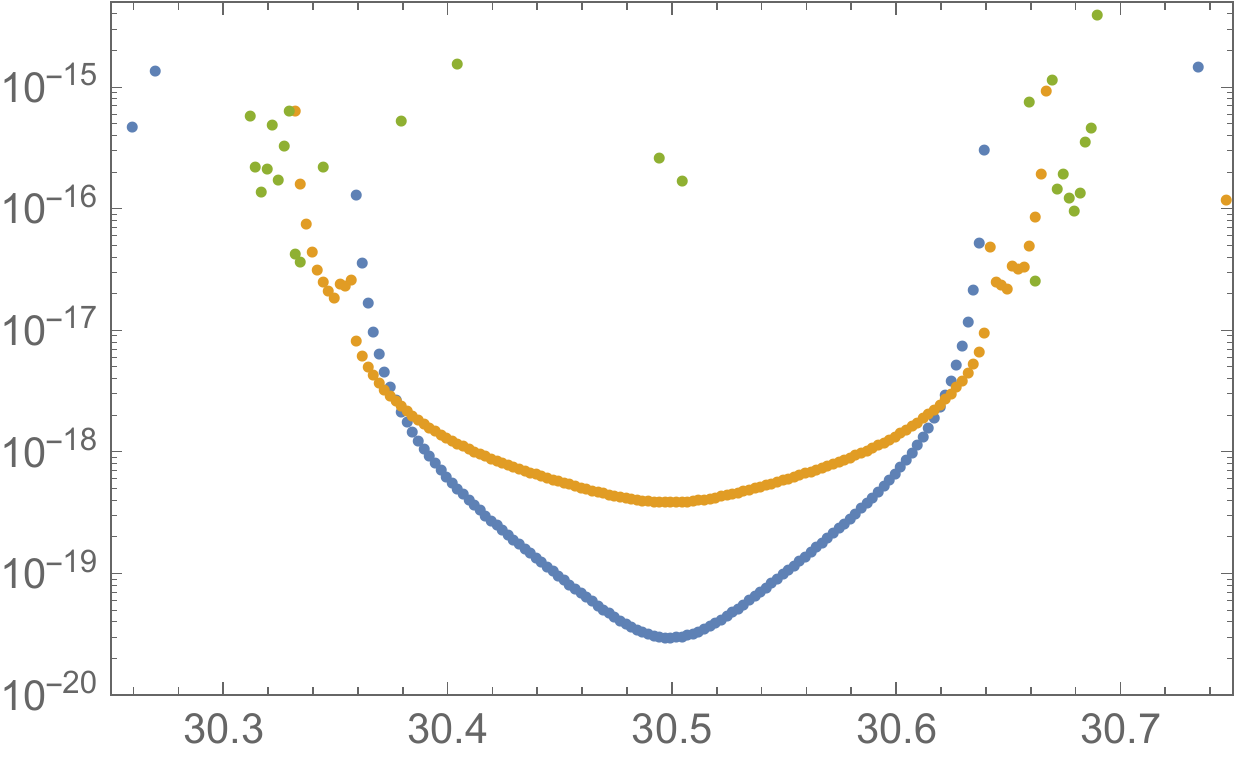}
\caption{The  left
  plot shows $(2\pi)^2 \Delta \bar \rho^{(2)}_i$, blue, orange, and green
  lines are for $i=1,2,3$ respectively, and the black dotted lines
  correspond  to
  (\ref{drho1},~\ref{drho1}).  The right plot shows
  $\delta \rho^{(n)}$; the blue, orange and green points correspond to
  $n=2,4,6$.  Only points with
  ${\rm Im} (\bar \rho^{(n)}) > 0.01 {\rm Re} (\Delta \bar
  \rho^{(n)})$ are shown. All plots are for $\hat \xi =10$.}
\label{F:vacuum} 
\end{subfigure}
\caption{}
\end{figure}

The left top plot in fig.~\ref{F:vacuum1} shows the exact critical
$k_n$ for which $\eps_n =1$ (using \eref{chiT} as the inflaton
background) for $n=1,..,4$ during a time interval $\delta T =1/2$;
this is compared with our approximation \eref{kn} . The approximation
is good (it becomes more accurate with larger $T$), except for parts close
to the critical points $T \is 3/8,1/2,5/8$ where either the sine or
the cosine goes to zero.

For the above estimate we only included the classical contribution to
the effective mass \eref{Meff}.  Taking into account the quantum
correction, even if larger than the tree-level term, will have
negligible effect.  As discussed in the previous subsection $\bar G$ is
nearly constant at late times, and the time-derivative of $\omega_k$
is still given by the time-derivative of the tree-level term.
Further, for the $q_0$ values of interest
$k_n^2 > M_{\rm eff}^2$, and one can use the approximation
$\omega_k \sim k$; the backreaction will not change this either.

If we compare the non-adiabatic modes with the tachyonic modes
\eref{ktach}, we see that $k_{\rm tach} <k_n$.  The tachyonic
modes give a an exponentially enhanced contribution to the Green function
and energy density, signifying particle production.  The larger the
enhancement, i.e., the larger $q_0$, the smaller the effect of the
different adiabatic subtractions.  However, since $k_n$ grows with
time, at sufficiently late times the vacuum dependence of $\bar G$ and
$\bar \rho$ always become substantial.  This is what we will discuss
next.

\subsection{Green function}

We now approximate the vacuum dependence of the Green function.  The
adiabatic frequency is very schematically of the form
\be
\( W_k^{(2m)}\)^2 = \omega_k^2 + f^{(2m)} ( \eps_1, \eps_2, ...,\eps_{2m}),
\ee
that is, $f^{(2m)}$ contains terms up to $2m$ derivatives of the
frequency $\omega_k$.  The highest derivative term dominates, and thus
$W^{(2m)} > W^{(2m-2)}$ for modes $k < k_{2m}$ for which
$\eps_{2m} > 1$.  Consider the difference between the Green functions of
the zeroth and $n$th order adiabatic vacuum. This is approximately
\begin{align}
\Delta \bar G^{(2m)} & \equiv \bar G^{(2m)}- \bar G^{(0)} = \int
  \frac{d^3k}{(2\pi)^3} \left(\frac{1}{2W_k^{(0)}}-
    \frac{1}{2W_k^{(2m)}}\right)
\approx 
\frac1{4\pi^2}\int^{k_{2m}} \dd k \,\frac{k^2}{W^{(0)}_k}
\nn \\
&\sim \frac1{8\pi^2} k_{2m}^2
\label{dG0}
\end{align}
In the last step we used that $W^{(0)}_k = \omega_k \sim k$, which is
a good approximation for the modes that dominate the
integral. Comparing with the exact result, we can fit the constant of
proportionality
\be
\Delta \bar G^{(2)} \approx \frac1{8\sqrt{2}\pi^2} k_2^2,\quad
\Delta \bar G^{(4)} \approx \frac1{16\pi^2} k_4^2,\quad
\Delta \bar G^{(6)} \approx \frac1{16\sqrt{2}\pi^2} k_{6}^2.
\label{dG}
\ee
The top right plot in figure \ref{F:vacuum1} shows the exact
$\Delta \bar G^{(n)}$ for $n=2,4,6$
and our approximation \eref{dG} over a time interval $\delta T =1/2$.  We only show points for which
the imaginary part of $\Delta \bar G^{(n)}$ is less than 1\%. That is, the modes
that dominate the integral are all non-tachyonic $W^{(n)}_k >0$ and
the adiabatic vacuum is defined for them, but we allow that a fraction
of the smaller $k$-modes (that give a subdominant contribution) are
tachyonic.  The vacuum dependence of the Green function grows with
time and for larger adiabatic order.  However, one cannot take the
order to infinity and claim that the vacuum dependence is arbitrarily
large, as for larger $n$ the frequency is tachyonic $(W_k^{(n)})^2 <0$ for
more $k$-modes and the Green function becomes increasingly less
well-defined. Indeed, the plot already shows that for $n=6$ the imaginary
part of the Green function exceeds 1\% during most of the time
interval, also at times where lower order vacua and $\bar G^{(2)},
\bar G^{(4)}$ are
still well defined.

The second observation is that the vacuum dependence $\Delta \bar G^{(n)}$
is minimized for $T \is 1/8,3/8$, in accordance with our estimate
\eref{kn} as $k_{2m}$ vanishes here. It seems that zooming in on
this time instant the vacuum dependence can be made arbitrarily
small. However, this is not the case, as in this limit more and more
$k$-modes become tachyonic, and at some point the Green function is no
longer well defined.  This still puts a lower bound on the vacuum
dependence.  Moreover, at these time instants the frequencies are
still non-adiabatic (only the even adiabaticity parameters vanish),
and this vanishing is more a coincidence rather than a consequence of
adiabaticity. Indeed, as we will see next, the vacuum dependence of
the energy density is minimized at different time instants, and one
cannot make both of them arbitrarily small at the same time.

\subsection{Energy density}

Define the vacuum dependence of the energy density via
\be
\Delta \bar \rho^{(n)} \equiv \bar \rho^{(n)}- \bar \rho^{(0)}, 
\ee
i.e. we compare the zeroth and $n$th order adiabatic vacuum.  To find an
analytic approximation it is useful to look at the different terms in
$\bar \rho_{\rm ad}$ separately
\begin{align}
 \bar{\rho}^{(n)}_{\rm ad}&=   -\frac{1}{4} \int \frac{d^3 k}{(2\pi)^3} 
 \biggl\{  \frac{1}{4}
  \frac{(W_k^{(n)'})^{2}}{(W^{(n)}_k)^3} + \left( W^{(n)}_k  + \frac{k^2}{W^{(n)}_k}\right)
+ \frac{ \hat\xi a^2 R}{W^{(n)}_k}
\biggr\}
\nn \\
& = \bar{\rho}^{(n)}_1 + \bar{\rho}^{(n)}_2
+ \bar{\rho}^{(n)}_3.
\label{drhoi}
\end{align}
For definiteness, concentrate on the $n=2$ vacuum; the results can be
generalized to higher order vacua.  The term $\Delta \bar \rho_3$ is
subdominant for $T \gtrsim \sqrt{q_0}$ and is neglected. The
term $\Delta \bar \rho_2$ is dominated by the modes $k \sim k_2$ for
which $W^{(0)} \sim W^{(2)}$, and can be
estimated by
\be
|\Delta \bar{\rho}^{(2)}_2|
\sim \frac1{4\pi^2} \int^{k_2} \dd k\, k^2 \times \frac{k}{2}
\sim  \frac{1}{64\pi^2}  k_2^4,
\label{drho1}
\ee
where the factor $1/2$ is matched to the numerical solution.  The term
$\Delta \bar \rho_1$ is also dominated by modes $k\sim k_2$, as the
integrand is peaked for these momenta.  We estimate
\be
|\Delta \bar{\rho}^{(2)}_1| \approx  \frac{1}{32\pi^2} 
 \int^{k_2}  k^2 \dd k \,  \frac{(W_k^{(2)'})^{2}}{(W^{(2)}_k)^3} 
\sim  \frac{(W_k^{(2)'})^{2}\big|_{k\sim k_2}}{32\pi^2 (3/2)^{3/2}} 
\sim \frac1{256\sqrt{6}} \frac{k_3^{10}}{k_2^{6}},
\label{drho2}
\ee
where once again the numerical factors are matched to the numerical
solution.  It should not surprise that $k_3$ enters this estimate,
as this term involves an extra time derivative and thus depends on the
3rd adiabaticity parameter.

The different contributions to $\Delta \bar \rho^{(2)}$ and the analytical
approximation (\ref{drho1},~\ref{drho2}) are shown in the bottom left plot in
Fig.~\ref{F:vacuum}. The total vacuum dependence of
$\Delta\bar \rho^{(n)}$ is shown in the bottom right plot for
$n=2,4,6$. Just as for the Green function, only the time instants are
shown where the imaginary contribution is less than 1\%.

Both approximations agree well with the numerical result away from the
critical points $T\is 3/8,1/2,5/8$.  However, in these limits
increasingly more modes become tachyonic, and the energy density is at
some point not well defined.  The vacuum dependence grows with the
order of the vacuum $n$, but just as for the Green function case one
cannot take arbitrary large $n$ as these vacua are tachyonic and not
defined during most of the time (some of the scattered points, mostly for
$\Delta \bar \rho^{(6)}$, are a numerical artefact as the integrand
behaves wildly for these points).

$\Delta \bar\rho^{(n)}$ is minimized for $T\is 1/2$ and blows up at the
other critical points $T\is 3/8,5/8$. 

In the following we will use $\bar G^{(2)}$ and
  $\bar \rho^{(2)}$ as a an estimate for the vacuum dependence.
Away from the critical points to a good approximation
\be
\Delta \bar G^{(2)} \sim \frac1{8\sqrt{2}\pi^2} k_2^2,
\quad
\Delta \bar \rho^{(2)} \sim {\rm max} \[ \frac{1}{64\pi^2} k_2^4,\, \frac1{256\sqrt{6}} \frac{k_3^{10}}{k_2^{6}}\].
\label{dG2}
\ee

\section{Vacuum stability}\label{s:vacstab}

In this section we discuss the implications of our results for the vacuum stability. In the next subsection we first discuss the criteria
for stability. In \ref{s:timescales} we analytically determine all the relevant
time scales, which (where possible) we compare with the numerical
results in \ref{s:numerics}.

\subsection{Criteria for stability}
\label{s:stability}

The qualitative form of the potential depends sensitively on the
Hubble scale.  If the Hubble scale is larger than the critical scale
$H > \mu_{\rm cr} \sim 10^{11}$GeV it follows from \eref{mu} that
$\lambda(\mu) <0$ for all Higgs field values $\phi < H$.  This is the
case at early times, for a sufficiently small number of inflaton
oscillations. Using \eref{RH} the critical time is given by\footnote{Here we
  used that the ratio $A_T/A_0$ is approximately constant for
  different $A_0$.} 
\be
T_{\rm crit}  \sim 25 \(\frac{A_0}{0.5}\)\( \frac{m_\chi}{10^{-5}}\)\( \frac{4.2 \times
  10^{-8}}{\mu_{\rm cr}}\).
\label{Tcrit}
\ee
with $A_0$ the initial inflaton amplitude. There are then two
possibilities for the vacuum to decay.

\begin{itemize}

\item The quantum corrections to the effective mass \eref{Meff}
  dominate before $T_{\rm crit}$. 

  As we have seen the Green function is nearly constant after a few
  oscillations. As a consequence, if the coupling is negative, the quantum correction to the effective Higgs mass is then always negative.  If it exceeds
  the (oscillating) tree-level contribution the vacuum is
  destabilized, since there is no barrier.  The vacuum can only remain stable if
  the following criterion is satisfied
 \be 
{\rm \bf condition} \;\;{\bf 1}: \qquad
6 \lambda \bar G^{\rm ren}(\tau) \lesssim  a^2 \hat \xi
R_{\text{max}}, 
\quad {\rm for} \;\; T < T_{\rm crit},
\label{condition1}
\ee
with $R_{\rm max}$ the maximum value of the Ricci scalar during an
inflaton oscillation.

\item
The energy density becomes larger than the potential barrier
separating the electroweak and true vacuum.

If the tree level term dominates the effective mass the potential is
an oscillating function with a barrier separating the two vacua. If
the quantum correction to the effective mass comes to dominate at late
times $T > T_{\rm crit}$, this only enhances the barrier, as the 
quantum correction to the effective mass is positive at
$\phi < \mu_{\rm cr}$.    

The tunnelling process is expected to be enhanced if the energy density
in the Higgs modes exceeds the height of barrier, potentially leading
to rapid decay.\footnote{Although a full calculation is needed to
  assert whether this is indeed the case. It should be noted that both the
  energy density and the potential are oscillating functions.}  In addition to the first criterion, we thus have a second criterion for vacuum stability:
\be
\hspace{-1.5cm}
{\rm \bf condition} \;\;{\bf 2}: \hspace{1.cm}
\bar\rho^{\rm ren} < \bar{V}_{\text{max}},\hspace{2.5cm}
\label{condition2}
\ee
with $\bar V_{\rm max}$ given in \eref{Vmax}.

\end{itemize}

In the literature, many different criteria for stability are used
\cite{rajantie2,ema,Kohri:2016wof,Kohri:2016qqv,Enqvist:2016mqj}.
Even though the details may differ, all agree that for large
non-minimal coupling $q_0 \gtrsim \O(5)$ particle production is
efficient and the electroweak vacuum is destabilized within a couple
of inflaton oscillations, since the stability condition \eref{condition1}
is violated.

For smaller couplings  particle production is initially
not efficient enough to destabilize the vacuum.  As the
classical background red shift faster than the quantum correction, the
energy density may become larger than the effective barrier of the
potential at late times, and \eref{condition2} is violated.  However,
as we will see, the vacuum dependence of the energy density grows with
time, and this makes it hard to make definite statements for small
$q_0 \sim \O(1)$.

\subsection{Time scales}
\label{s:timescales}

In this section we discuss the relevant time scales, starting with
some effects we have neglected (inflaton and Higgs decay), and ending
with estimates for when the Green function and energy density become
large, when the vacuum dependence becomes important, and when the
stability criteria are violated.

\subsubsection{Inflaton decay}

Even though the inflaton decay rate is model dependent, we can make
some general statements.

The inflaton oscillates until decay.  If it decays via very efficient
preheating (into a field other than the Higgs field), the total number
of oscillations are $T \lesssim 10^3-10^4$ \cite{Kofman2}.
Perturbative decay is slower.  The inflaton cannot be coupled too
strongly to other fields, as otherwise loop corrections would spoil
the flatness of the potential. This gives a bound $T \gtrsim 10^6$
corresponding to a maximum reheating temperature
$T_R \lesssim 10^{12}$\GeV ; but the number of oscillations can be
much larger, for example if the coupling is of gravitational strength
and $\Gamma_\chi \sim m_\chi^3/\mpl^2$, decay happens after
$T \sim 10^9$ oscillations, with a maximum reheating temperature
$T_R \lesssim 7 \times 10^{10}$GeV.

After inflaton decay the universe is dominated by a relativistic
fluid, and $R \approx 0$.  The classical contribution to the Higgs
mass vanishes.  However, already before complete decay, there are
temperature corrections to both the Higgs mass and the energy density,
just like the non-thermal preheating corrections discussed in this
paper.  The net effect is that the thermal corrections generically
stabilize the vacuum --- see \cite{Espinosa:2007qp,higgstory} for the
exact bounds.  For non-perturbative decay, there is an intermediate
non-thermal stage before thermal equilibrium is reached; during this
stage the Ricci scalar is non-vanishing. Although a dedicated study
will be needed to determine the exact bounds in this case, it is
likely that perturbative inflaton decay also helps to stabilize the
vacuum.

In the next subsections we will discuss the critical $q_0$ value below
which the electroweak vacuum is stable, assuming inflaton decay is
late and only happens after all relevant time-scales.  Since inflaton
decay generically stabilizes the electroweak vacuum, fast decay
can only raise the critical $q_0$ value.

\subsubsection{Higgs decay}

In coordinate time the Higgs decays when $\Gamma \sim H$, provided
decay is kinematically allowed.  Translated to conformal time, the
condition for decay is
\be
a H \sim \alpha M_{\rm eff},
\label{Hdecay}
\ee
with $\alpha = y^2/(4\pi)$ and $y$ a yukawa/gauge coupling.  For the
top quark and gauge couplings $\alpha\sim (1-5) \times 10^{-2}$ at
scales $\mu_{\rm cr}$, and for the bottom quark, tau and Higgs self coupling
$\alpha \sim 10^{-4} -10^{-5}$. The fermion/gauge boson mass in
conformal time, the equivalent of $M_{\rm eff}$ for the Higgs field
\eref{Meff}, only has a contribution from the Higgs fluctuations
$M_y \sim y^2 \bar G$.  Averaged over one oscillation
$H^2 \sim R \sim A^2 m_\chi^2/a^3$, as follows from \eref{RH}.

Consider first the case that the tree-level term dominates the Higgs
mass $M_{\rm eff}$. The Higgs is much heavier than the SM particles,
and decay into top quarks and gauge bosons is kinematically possible.
Decay only happens for $\sqrt{\xi} \alpha >1$, so only for large couplings
$\xi \gtrsim 10^4$.

In the opposite limit that quantum corrections dominate the Higgs
mass, decay is kinematically allowed if $M_{\rm eff} > 2M_y$. Since
both the mass of the Higgs and the mass of the decay products scale with
$\sqrt{\bar G}$, decay into top quark and EW gauge
bosons is impossible.  Decay into lighter fields is still allowed, in particular
decay into b-quarks, tau-leptons and Higgs will dominate. The Higgs decays when
\be
 \frac{\lambda \bar G}{a^2 \xi R} \gtrsim \frac{1}{\alpha^2 \xi}
\gtrsim \frac{10^8}{\xi},
\ee
where we used $\alpha \lesssim 10^{-4}$ in the last step.

Perturbative Higgs decay only happens at late times for the $\xi$
couplings of interest, and can be neglected for our purposes.

\subsubsection{Large Green function corrections}

At late times the maximum value of the classical mass during an
oscillation can be approximated by
\be
M_{\rm eff}^2|_{\rm class} = a^2\hat \xi R|_{\rm max} = 2 A_T^2
a_T^2 m_\chi^2 \hat \xi  T^{-\frac23}
+\O(T^{-\frac52}),
\ee
where we used \eref{RH}. The quantum contribution to the mass
$M_{\rm eff}^2|_{\rm quant} = 6 |\lambda| \bar G $ becomes of the same
order as the classical contribution at time
\be
T_{\rm G} \approx 5 \times 10^5 \(\frac{10^{-2}}{\lambda}\)^{3/2}
\e^{-\frac32 c\sqrt{q_0}},
\label{TG}
\ee
where we used \eref{Gfit}, and $c$ is given in \eref{cval}. We find 
$T_G \approx \{4 \times 10^3,\, 7 \times 10^2,\, 13, \, 0.2 \}$ for
$q_0 =\{1,\,2,\,5,\,10\}$. Preheating is efficient and the vacuum is
destabilized if the quantum correction dominates already when
$H > \mu _{\rm cr}$, see our first stability condition
\eref{condition1}.  This requires $T_G < T_{\rm crit}$, with $T_{\rm
  crit}$ given in \eref{Tcrit}:
\be
T_{\rm G} \lesssim T_{\rm crit}
\quad \Rightarrow \quad
q_0 \gtrsim q_0^{\rm crit} = 4.5 \;\;({\rm or} \; \xi \gtrsim 24).
\label{qcrit}
\ee
This is reasonably close to the value $q_0 ^\text{crit} \approx 3.7$ that we find numerically.

\subsubsection{Large energy density}

For $T > T_{\rm crit}$ the potential always has a barrier; decay may
still be possible if the energy density exceeds the maximum of the
potential, see our second stability condition \eref{condition2}.
The interesting parameter region is $q_0 <q_0^{\rm crit}$.  In that case
the energy density becomes large at a time
\be
T_\rho \sim 
2 \times 10^{3} \e^{-\frac34 c \sqrt{q_0}} 
\(\frac{10^{-2}}{|\lambda|}\)^{3/4},
\label{Trho}
\ee where we used approximation (\ref{rhofit}).  The energy
density becomes large before the quantum correction to the Green
function becomes large, that is $T_\rho < T_G$, and the above
time-scale was derived using the tree-level potential.  Taking
$|\lambda| = 10^{-2}$,
$T_\rho \approx \{2 \times 10^2,\, 55,\, 6, \,0.8\}$ for
$q_0 =\{1,\,2,\,5,\,10\}$.

\subsubsection{Vacuum dependence large}

We focus on the difference between the 0th and 2nd order adiabatic
vacuum, as a measure of the vacuum dependence of the results.  The
vacuum dependence becomes of the same order as the Green function itself
if
\be
\frac{\Delta \bar G^{(2)}}{\bar G^{(0)}} >1
\quad \Rightarrow 
\quad T > T_{\Delta G} = 6 \times 10^{-3} q_0^{3/2}\e^{3c \sqrt{q_0}} ,
\label{vacdepprop}\ee
where we took $\cos(4\pi T) \sim 1$.  This gives
$T_{\Delta G} =\{ 80,\, 10^5,\, 10^8\} $ for $q_0=\{1,\,2,\, 5\}$.
$T_{\Delta G} <1$ for $q_0 <1$.  The Green function enters the first
stability criterion \eref{condition1} for effective preheating, which
is valid if the vacuum decays within the first $~25$ oscillations and
$q_0 > q_0^{\rm crit} =4.5$. It follows from \eref{vacdepprop} that
the vacuum dependence of the results is small in this case.

If we compute the energy density at the instants where the vacuum
dependence is minimized $T \hat = \frac{1}{2}$, we have
$\Delta \bar{\rho}^{(2)} = \frac{1}{64\pi^2} k_2^4$ and the vacuum
dependence of the energy density becomes sizeable when
\be
\frac{\Delta \bar \rho^{(2)}}{\bar \rho^{(0)}} >1
\quad \Rightarrow \quad 
T >  T_{\Delta \rho} = 0.21 e^{\frac{3}{2} c \sqrt{q_0}} q_0^{2}.
\label{Tdrho}
\ee
This gives $T_{\Delta \rho} \approx \{40,\, 10^3 ,\, 3 \times 10^5\} $ for
$q_0=\{1,\,2,\, 5\}$. The energy density enters the second stability
criterion \eref{condition2} for $q_0 < q_0^{\rm crit}$.  Demanding that
\be 
T_{\rho} < T_{\Delta \rho} \quad \Rightarrow \quad
q_0 \gtrsim 1.3 \quad ({\rm or} \; \xi \gtrsim 7).
\ee
However, $V_{\rm max}$, $\bar{\rho}$ and $\Delta \bar{\rho}$ are all
oscillating functions. We can not expect that an interpolation between
the points where the vacuum dependence of $\bar \rho$ is minimized
gives a proper description of $\bar \rho$ during the whole
oscillation, or of the oscillation averaged $\bar \rho$. For an
accurate determination of the tunneling rate to the true vacuum,
information at other time instants may be needed.

Moving slightly away from $T \hat = \frac{1}{2}$, the vacuum
dependence is given by
$\Delta \bar{\rho} ^ {2} = \frac{1}{256 \sqrt{6}}
\frac{k_3^{10}}{k_2^6} $
and $T_{\Delta \rho}$ becomes much smaller. Taking
$T \hat= \frac{1}{2} + 0.05$ as an example, we get:
\be
\frac{\Delta \bar \rho^{(2)}}{\bar \rho^{(0)}} >1
\quad \Rightarrow \quad 
T >  T_{\Delta \rho} = 0.27 e^{\frac{3}{5} c \sqrt{q_0}} q_0^{9/10},
\ee
which gives $T_{\Delta \rho} \approx \{2,\, 10  ,\, 10^2\} $ for
$q_0=\{1,\,2,\, 5\}$. Demanding again that $T_\rho < T_{\Delta\rho}$
gives the stronger bound
\be
T_{\rho} < T_{\Delta \rho} \quad \Rightarrow \quad
  q_0 \gtrsim 2.8  \quad ({\rm or} \; \xi \gtrsim 15).
\ee
 
Numerically we find $q_0 \sim 0.9$ (or $\xi \sim 5$) when we look at
the vacuum dependence at $T\hat= \frac{1}{2}$, and $q_0 \sim 2.2$ (or
$\xi \sim 12$) for $T \hat= \frac{1}{2} + 0.05$.

\subsection{Numerical results}
\label{s:numerics}

In this section we present the results of our numerical calculation,
which serves as a check for the analytical results discussed in the
previous sections. We focus on the calculation of the Higgs Green
function and energy density, and in particular on their vacuum
dependence by calculating them in both the zeroth and second order
vacuum \eref{secondorder}.

\subsubsection{Numerical implementation}\label{s:implementation}

Since our aim is not so much on getting the most accurate results, but
rather on investigating the vacuum dependence, we have made some
simplifying assumptions to speed up the numerics.

First of all, we neglect the Higgs mass term and quartic Higgs
coupling.  The electroweak scale mass is negligible during
preheating. The quartic coupling only becomes important near the
maximum of the potential, but can be neglected for smaller field
values.  Since we are only interested in the question whether the EW
vacuum is destabilized, and not in the exact dynamics of the Higgs
after crossing the barrier, we can neglect this interaction.  This is
supported by numerical simulations in the literature (see
e.g. \cite{ema}).

Secondly, we neglect backreaction.  It can be checked that the
backreaction on the inflaton equation of motion of the produced
particles is small.  The backreaction for the Higgs mode equation
becomes important once the barrier disappears and $M^2_{\rm eff}$
becomes negative \eref{Meff}. At this point the vacuum is already
destabilized.

Thirdly, we neglect the interaction of the Higgs with all other SM
particles and also other inflaton interactions (which should be there
for complete decay of the inflaton). As will be discussed in
\ref{s:timescales}, Higgs decay happens on large time-scales and can
be neglected; inflaton decay is model dependent, but is expected to
only increase stability.

Since the inflaton background is formulated in coordinate time
\eref{eominf}, we solve for the Higgs modes in coordinate time as
well.  The mode functions satisfy the differential equation:
\be
  \ddot{U}_k + H \dot{U}_k + \frac{\omega_k^2}{a^2} U_k 
= 0, \qquad \omega_k^2 = k^2 + a^2(\xi - \frac{1}{6})R.
\ee
As suggested in \cite{Baacke:1997rs}, we write the mode functions as 
\be
  U_k = e^{-i\omega_{k,0}\tau}(1+h_k), \qquad h_k(t_0) 
= \dot{h}_k(t_0) = 0.\label{Ukhk}
\ee 
These $h_k$-functions satisfy the mode equation:
\be
  \ddot{h}_k + \(H - \frac{2 i \omega_{k,0}}{a} \)\dot{h}_k
 = -\(\xi - \frac{1}{6}\)\(R -\frac{a^2(t_0)}{a^2}R(t_0)\)(1+ h_k).
\ee
The mode equations are solved numerically in Mathematica. The
$h_k$-modes are numerically more stable for large $k$ than the
$U_k$-modes. 

We are interested in the quantities $\bar \rho^{(n)}$ and $\bar G^{(n)}$, which are obtained by
integrating over $U_k$-modes. In order to keep the computation time
manageable, the integral is replaced by a sum. Typically we compute
$U_k$ and $\dot{U}_k$ for 100 values of $k$ lying on a logarithmic
scale between $10^{-7}$ and $10^{-3}$. The mode functions do not vary
significantly with $k$ for $k$ as small as $10^{-7}$. Furthermore, the
contribution for small $k$ is suppressed by the factor $k^2$ from the
Jacobian, so our lower boundary is effectively zero. The contribution
of the large $k$ modes decreases exponentially beyond a certain cutoff
(much smaller than $k = 10^{-3}$), due to the adiabatic
renormalization, and these modes can be neglected.

Since we use the adiabatic approximation for renormalization, our
expressions for $\bar\rho$ and $\bar G$ are only properly defined when
$W_k^2(t) >0$. In order to have a continuous graph, we either
interpolate between values of $T$ for which $W_k^2(t) >0$ when we focus
on large time scales, or, when we focus on shorter time
scales, we take the absolute value of the renormalization terms in
equation \eref{Gren}. We sample

\subsubsection{Results}

\begin{figure}
\centering
\begin{subfigure}{1\textwidth}
\centering
  \includegraphics[width = 0.46 \linewidth, bb = 21 3 356 240]{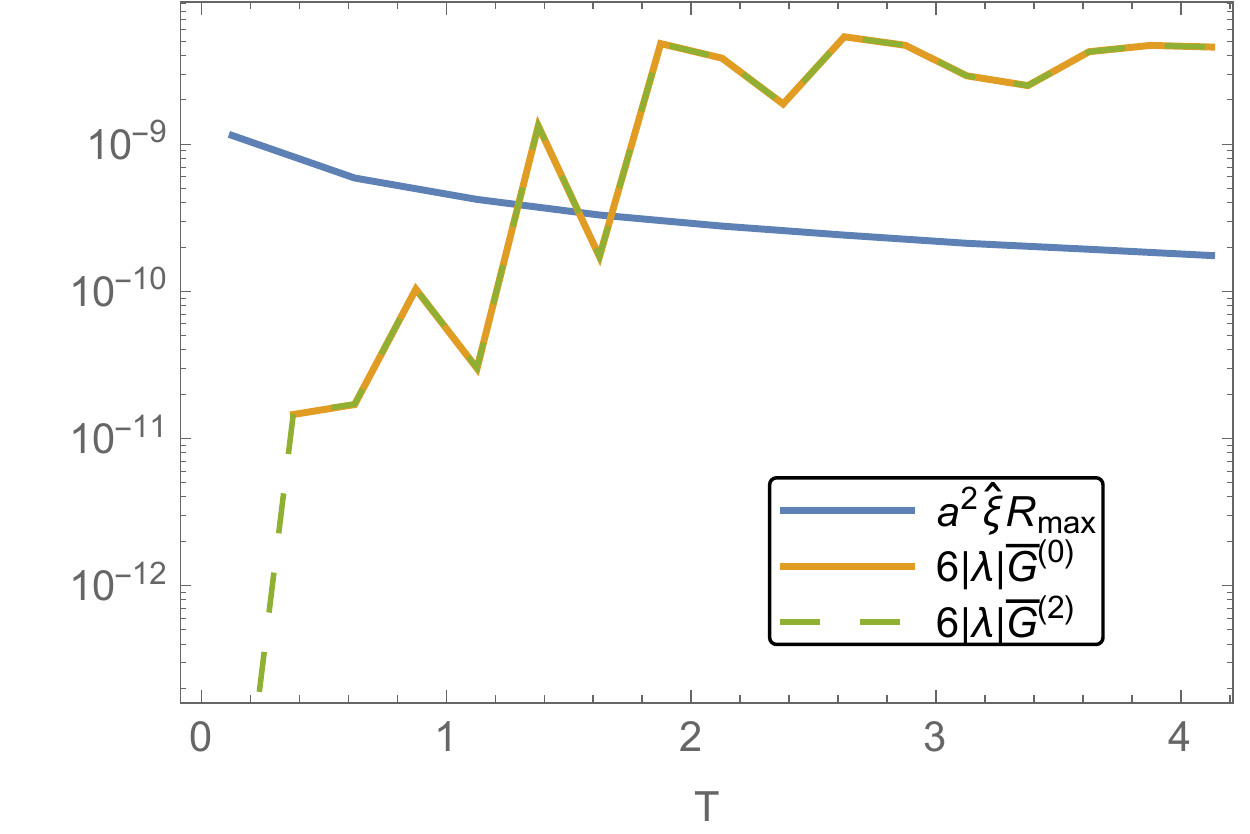}
\includegraphics[width = 0.5 \linewidth, bb = 20 3 359 224]{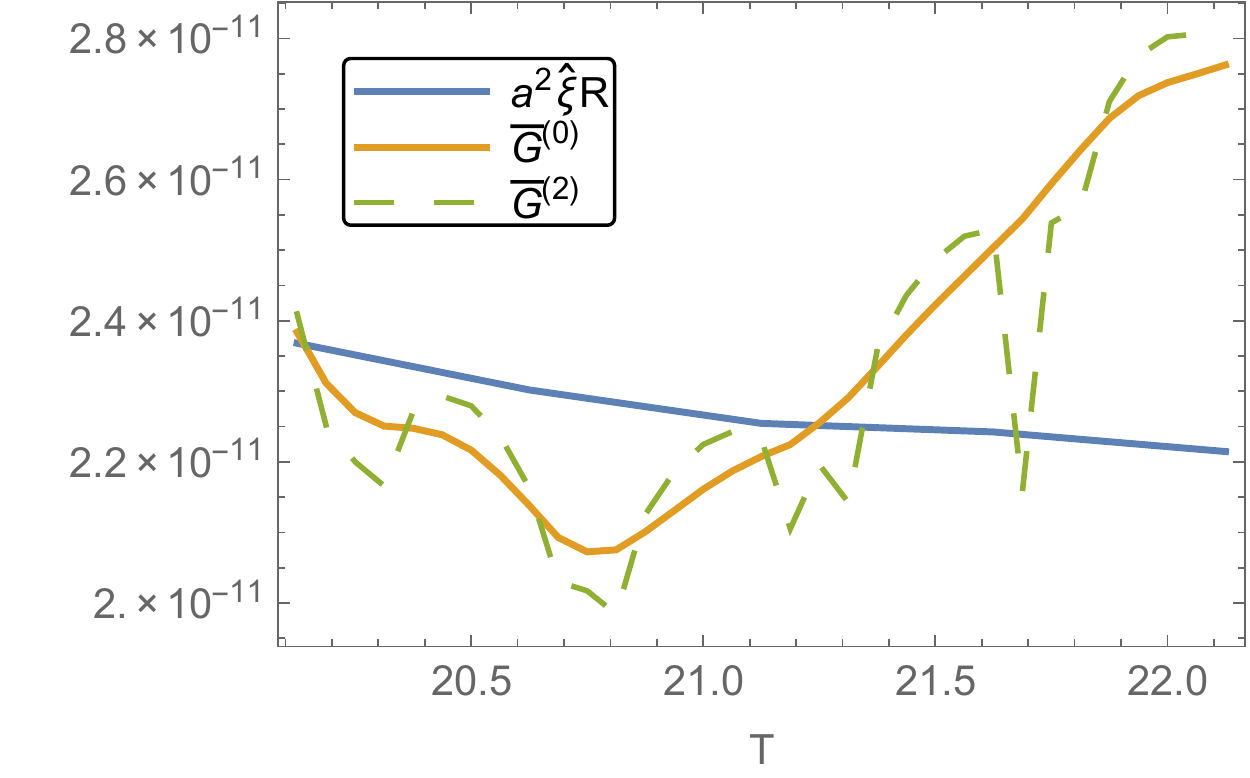}
  \caption{Left: $q_0 =9.3$. Tree-level contribution $a^2 \hat \xi R$ (blue)
    and one-loop contribution $6|\lambda| \bar G$ (orange and green)
    to the effective Higgs mass for the first few oscillations. The
    $a^2\hat \xi R$-term oscillates in reality, but in this graph we
    show an interpolation between the maxima. The solid orange line
    shows the result for $\bar{G}$ in the zeroth order adiabatic
    vacuum, the dashed green line shows the second order result. In
    order to have a continuous plot, we take the absolute value of the
    renormalization terms in equation \eref{Gren}. $\bar{G}$ is sampled with $\delta T = 1/4$
Right: $q_0 = 3.7$. (Interpolation between the maxima of)
    tree-level contribution $a^2 \hat\xi R$ (blue) and one-loop
    contribution $6|\lambda| \bar G$ (orange and green) to the
    effective Higgs mass at $T \approx 21$. The solid orange line
    shows the result for $\bar{G}$ in the zeroth order adiabatic
    vacuum, the dashed green line shows the second order result. In
    order to have a continuous plot, we take the absolute value of the
    renormalization terms in equation \eref{Gren}. $\bar{G}$ is sampled with $\delta T = 1/16$.}
\label{fig:xi}
\end{subfigure} \\
\begin{subfigure}{1\textwidth}
\centering
\vspace{1cm}
  \includegraphics[width = 0.46 \linewidth, bb = 20 3 356 233]{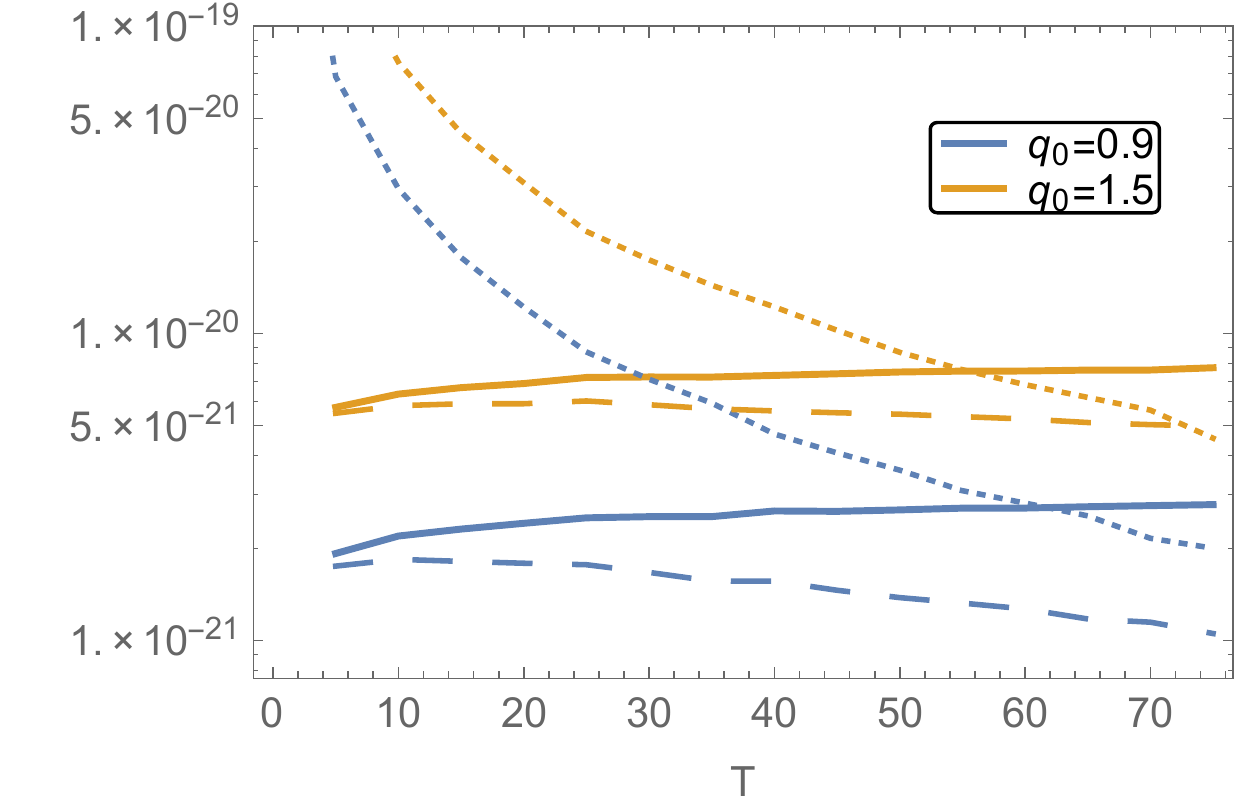}
  \includegraphics[width = 0.46 \linewidth, bb = 20 3 419 268]{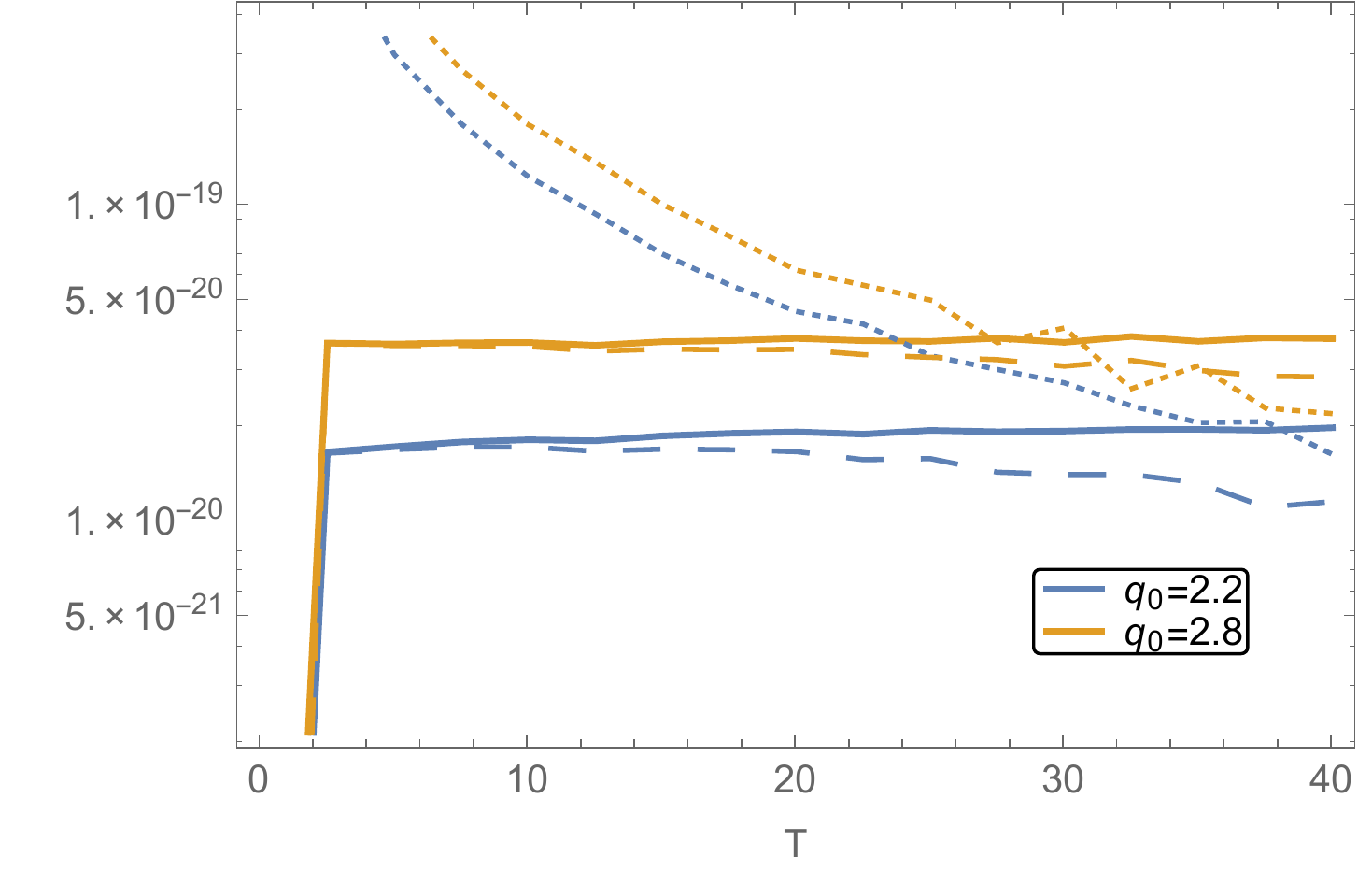}
  \caption{Interpolation of the maxima of $\bar\rho$ and
    $\bar V_\text{max}$ as a function of time. The solid line shows
    $\bar \rho^{(0)}$, the dashed line shows $\bar \rho^{(2)}$, the
    dotted line shows $\bar V_\text{max}$ in the zeroth order vacuum
    (the second order result looks identical). Left: $q_0 = 0.9$ (blue) and $q_0 = 1.5$ (orange).  $\bar\rho$ and $\bar V_\text{max}$ are sampled with $\delta T = 5$, starting at $T = 0$, such that the vacuum dependence is minimized. Right: $q_0 = 2.2$ (blue) and $q_0 = 2.8$ (orange). $\bar\rho$ and $\bar V_\text{max}$ are sampled with $\delta T = 2.5$, starting at $T = 0.05$.}
\label{fig:RhoVmax}
\end{subfigure}
\caption{}
\end{figure}

\paragraph{Large $q_0$, immediate decay of the Higgs}

Preheating is efficient for $q_0 \gg 1$, or equivalently $\xi \gg 1$,
and leads to decay of the electroweak vacuum. We show the results for
$q_0 = 9.3$ ($\xi =50$) in figure \ref{fig:xi}.  We
plotted both the oscillating tree-level contribution to the effective
Higgs mass and the negative one-loop contribution. Within one inflaton
oscillation the latter dominates and the first criterion for stability
\eref{condition1} is violated: the potential does not have a maximum
anymore and the Higgs will inevitably roll down to the true minimum.
It should be noted that once that happens, our results are inaccurate,
since it is no longer justified to neglect the quartic term from the
potential.  We tested for vacuum dependence, by calculating the
quantum correction in both the zeroth and second order adiabatic
vacuum \eref{secondorder}.  Only during the first half of the first
oscillation we can see a difference between the zeroth and second
order adiabatic renormalization, but the final result is nearly vacuum
independent, as expected from equation (\ref{vacdepprop}).

\paragraph{Large-$T$ behavior of $\bar{G}$ and $\bar{\rho}$.}
For smaller values of $q_0$, preheating is less efficient, and the
initial production of Higgs modes is not enough to let the Higgs decay
within the first few oscillations.  

As was shown in figure \ref{fig:largeT}, the Green function and energy
density remain approximately constant after the initial amplification.
Due to the different scaling of the classical and one-loop
contribution, the first stability criterion might still be violated
before $H$ becomes smaller than the instability scale $\mu_\text{cr}$
even if $q_0$ was not large enough to lead to immediate Higgs decay.

Figure \ref{fig:xi} shows the tree level and one-loop mass entering
the first stability criterion (\ref{condition1}) at $T\approx 21$ for
$q_0 = 3.7$ ($\xi =20$). The graph shows that the one-loop
contribution indeed becomes larger than the tree-level contribution,
resulting in decay of the Higgs. The difference between $\bar{G}$ in
the two vacua is approximately 50 times smaller than $\bar{G}$
itself. Since the Hubble constant becomes smaller than $\mu_\text{cr}$
around $T = 25$, the first criterion for stability will not be
violated for values of $q_0 \lesssim 3.7$. This value of $q_0$ is
reasonably close to the estimated value of $q_0^\text{crit} = 4.5$.

\paragraph{Comparison of $\bar V_\text{max}$ and $\bar\rho$ for $q_0 < 3.7$.}

For $q_0 < 3.7$, the Green function contribution does not become
dominant before $T = 25$, so we need to look at the second stability
criterion \eref{condition2} to determine the fate of the Higgs.

The left plot of figure \ref{fig:RhoVmax} shows a comparison between
$\bar{\rho}^{(0)}$, $\bar{\rho}^{(2)}$ and $\bar V_\text{max}$ for
$q_0 = 0.9, 1.5$ ($\xi = 5, 8$) for values $T \hat = \frac12$ where
the vacuum dependence is minimized. The computation of
$\bar V_\text{max}$ is done in the zeroth order vacuum, but the vacuum
dependence of $\bar V_\text{max}$ is in any case subdominant. For
$q_0 = 1.5$, $\bar{\rho}$ becomes comparable to $\bar V_\text{max}$
when the vacuum dependence is still rather small, implying that there
is a possibility of vacuum decay (remember though that $\bar{\rho}$
oscillates). For $q_0 = 0.9$, $\Delta\bar\rho ^{(2)}$ becomes
comparable to $\bar\rho^{0}$ at the same time when
$\bar{\rho}^{(0)} = \bar V_\text{max}$. As a consequence, we can not
determine whether the Higgs decays. Our critical value of $q_0$ agrees
reasonably well with the estimated value of $q_0 \approx 1.3$. We
conclude that for $q_0 \lesssim 0.9$ (or $\xi \lesssim 12$) we can not
draw vacuum-independent conclusions about the stability of the Higgs.

The right graph of figure \ref{fig:RhoVmax} gives an even stricter
bound on the value of $q_0$ for which vacuum-independent statements
can be made. $\bar{\rho}$ and $\bar V_\text{max}$ are determined at
$T \hat=\frac12+ 0.05$. We then find that no vacuum dependent
statements can be made for $q_0 \lesssim 2.2$.

\section{Conclusion}

In this work we studied the stability of the Higgs field during
preheating in the presence of a non-minimal coupling between the Higgs
field and the Ricci scalar. After inflation the inflaton oscillates in
its potential and the effective Higgs mass squared oscillates between
positive and negative values.  Higgs modes with momentum $k$ such that
$\omega_k^2 <0$ during part of the inflaton oscillation are produced
in a process called tachyonic preheating; in addition non-adiabatic
modes are produced n a process called parametric resonance.  The
produced Higgs modes contribute to the effective potential and the
energy density.

Vacuum decay can now occur in two ways.  First, if preheating is very
efficient it leads to explosive particle production and within a few
oscillations the effective Higgs mass is dominated by the one-loop
corrections. At early times the Hubble exceeds the critical
renormalization scale and the Higgs coupling is negative
$\lambda (\mu \sim H) < 0$, and consequently the effective Higgs mass
is negative. The Higgs will inevitably roll down to the true minimum.
Secondly, if preheating is less efficient, the one-loop contribution may
still dominate at late times as the background term red shifts away
faster.  At late times and for smaller Hubble scale, the effective
potential will always have a barrier separating the true and false
vacuum. However, the energy density may become comparable or larger
than the maximum of the potential, making tunnelling to the true vacuum
likely.  

Particle production is dominated by the initial times, where the Higgs
mass becomes most negative.  This allows to approximate the Green
function and energy density by the contribution during the first
oscillation, as it remains (nearly) constant afterwards.  We presented
the semi-analytic approximations in section \ref{s:analytic}.  We compared
the analytical estimate of the propagator and energy density with the
numerical values for values of the efficiency parameter $q_0$ between
$0.16$ and $9.3$ ($\xi$ between 1 and 50) and found an agreement up to
a factor of 2 for most values of $q_0$.

The results are defined with respect to the $n$th order adiabatic
vacuum.  The adiabatic vacuum is a good vacuum for high momentum
modes, for which the adiabaticity parameters defined in
\eref{adiabaticity} are small at least during part of the inflaton
oscillation.  However, for smaller momenta, adiabaticity is always
violated. We found that the contribution of the non-adiabatic momenta
to the Green function and/or energy density grows with time, and thus
the vacuum dependence grows with time.  Since different choices of
vacua correspond to different renormalization conditions, there is
thus an uncertainty in the values of the renormalized couplings ---
and in the definition of the critical coupling for vacuum stability
--- which can only be fixed (in theory) by a measurement during
preheating.

Our main results are that for $q_0 \gtrsim 3.7$ preheating is efficient
and the vacuum decays at early times when $H > \mu_{\rm cr}$. This is
in agreement with results in the literature
\cite{rajantie2,ema,Kohri:2016wof,Kohri:2016qqv,Enqvist:2016mqj}. Since
decay is rapid, vacuum effects are negligible in this case.

For $0.9.\lesssim q_0 \lesssim 3.7$ the Higgs potential always has a
barrier, but the energy density becomes of the order of the barrier
height before the results are swamped by uncertainties due to the
vacuum choice.  

For $q_0 \lesssim 0.9$ however, no definite statements can be made. This bound becomes stricter if we compare $\bar \rho$ and $\bar V_{\text{max}}$ at time instants where the vacuum dependence of $\bar \rho$ is not minimized. When
the energy density becomes large enough for the vacuum to be in peril,
the vacuum dependence is already large (the difference in energy
density for different vacua is larger than the energy density itself
calculated in the zeroth order vacuum).  As a consequence there is a
large uncertainty in the definition of the effective couplings, and
thus any attempt to determine a critical coupling for stability is
hampered by that.  Note that this is arguably the most interesting
part of parameter space, as order one couplings and initial conditions
such as in chaotic or Starobinsky inflation lead to $q_0 = \O(1)$.

\section*{Acknowledgments}

The authors are funded by the Dutch Organisation for Scientific
Research (NWO).  We thank Jan-Willem van Holten and Jacopo Fumagalli
for illuminating discussions.



\begin{thebibliography}{9}
  
  
   \bibitem{disc1} 
  F.~Bezrukov, M.~Y.~.Kalmykov, B.~A.~Kniehl and M.~Shaposhnikov,
  JHEP {\bf 1210}, 140 (2012)
  [arXiv:1205.2893 [hep-ph]].
  
  \bibitem{disc2} 
  G.~Degrassi, S.~Di Vita, J.~Elias-Miro, J.~R.~Espinosa, G.~F.~Giudice, G.~Isidori and A.~Strumia,
  JHEP {\bf 1208}, 098 (2012)
  [arXiv:1205.6497 [hep-ph]].
 
 \bibitem{branchina} 
  V.~Branchina and E.~Messina,
  Phys.\ Rev.\ Lett.\  {\bf 111}, 241801 (2013)
  [arXiv:1307.5193 [hep-ph]].

\bibitem{branchina2}
  V.~Branchina and E.~Messina,
  arXiv:1507.08812 [hep-ph].

\bibitem{archil} 
 A.~Kobakhidze and A.~Spencer-Smith,
  arXiv:1404.4709 [hep-ph].
  

\bibitem{alexss} 
  A.~Spencer-Smith,
  arXiv:1405.1975 [hep-ph].

\bibitem{kniehl}
  A.~V.~Bednyakov, B.~A.~Kniehl, A.~F.~Pikelner and O.~L.~Veretin,
  arXiv:1507.08833 [hep-ph].
  
\bibitem{Espinosa:2007qp}
  J.~R.~Espinosa, G.~F.~Giudice and A.~Riotto,
  JCAP {\bf 0805} (2008) 002
  doi:10.1088/1475-7516/2008/05/002
  [arXiv:0710.2484 [hep-ph]].

\bibitem{Enqvist:2014bua}
  K.~Enqvist, T.~Meriniemi and S.~Nurmi,
  JCAP {\bf 1407} (2014) 025
  doi:10.1088/1475-7516/2014/07/025
  [arXiv:1404.3699 [hep-ph]].

\bibitem{Fairbairn:2014zia}
  M.~Fairbairn and R.~Hogan,
  Phys.\ Rev.\ Lett.\  {\bf 112} (2014) 201801
  doi:10.1103/PhysRevLett.112.201801
  [arXiv:1403.6786 [hep-ph]].

\bibitem{Kobakhidze:2013tn}
  A.~Kobakhidze and A.~Spencer-Smith,
  Phys.\ Lett.\ B {\bf 722} (2013) 130
  doi:10.1016/j.physletb.2013.04.013
  [arXiv:1301.2846 [hep-ph]].

\bibitem{Hook:2014uia}
  A.~Hook, J.~Kearney, B.~Shakya and K.~M.~Zurek,
  JHEP {\bf 1501} (2015) 061
  doi:10.1007/JHEP01(2015)061
  [arXiv:1404.5953 [hep-ph]].

\bibitem{rajantie1}
  M.~Herranen, T.~Markkanen, S.~Nurmi and A.~Rajantie,
  Phys.\ Rev.\ Lett.\  {\bf 113} (2014) no.21,  211102
  doi:10.1103/PhysRevLett.113.211102
  [arXiv:1407.3141 [hep-ph]].


\bibitem{higgstory}
  J.~R.~Espinosa, G.~F.~Giudice, E.~Morgante, A.~Riotto, L.~Senatore, A.~Strumia and N.~Tetradis,
  JHEP {\bf 1509} (2015) 174
  doi:10.1007/JHEP09(2015)174
  [arXiv:1505.04825 [hep-ph]].
  



\bibitem{Kamada:2014ufa}
 K.~Kamada,
  Phys.\ Lett.\ B {\bf 742} (2015) 126
  doi:10.1016/j.physletb.2015.01.024
  [arXiv:1409.5078 [hep-ph]].



\bibitem{Kofman1}
  L.~Kofman, A.~D.~Linde and A.~A.~Starobinsky,
  Phys.\ Rev.\ Lett.\  {\bf 73} (1994) 3195
  doi:10.1103/PhysRevLett.73.3195
  [hep-th/9405187].

\bibitem{Kofman2}
  L.~Kofman, A.~D.~Linde and A.~A.~Starobinsky,
  Phys.\ Rev.\ D {\bf 56} (1997) 3258
  doi:10.1103/PhysRevD.56.3258
  [hep-ph/9704452].

\bibitem{Shtanov:1994ce}
  Y.~Shtanov, J.~H.~Traschen and R.~H.~Brandenberger,
  Phys.\ Rev.\ D {\bf 51} (1995) 5438
  doi:10.1103/PhysRevD.51.5438
  [hep-ph/9407247].


\bibitem{Felder:2000hj}
  G.~N.~Felder, J.~Garcia-Bellido, P.~B.~Greene, L.~Kofman, A.~D.~Linde and I.~Tkachev,
  Phys.\ Rev.\ Lett.\  {\bf 87} (2001) 011601
  doi:10.1103/PhysRevLett.87.011601
  [hep-ph/0012142].

\bibitem{Dufaux}
  J.~F.~Dufaux, G.~N.~Felder, L.~Kofman, M.~Peloso and D.~Podolsky,
  JCAP {\bf 0607} (2006) 006
  doi:10.1088/1475-7516/2006/07/006
  [hep-ph/0602144].


\bibitem{Tsujikawa:1999jh}
  S.~Tsujikawa, K.~i.~Maeda and T.~Torii,
  Phys.\ Rev.\ D {\bf 60} (1999) 063515
  doi:10.1103/PhysRevD.60.063515
  [hep-ph/9901306].

\bibitem{Bassett:1997az}
  B.~A.~Bassett and S.~Liberati,
  Phys.\ Rev.\ D {\bf 58} (1998) 021302
   Erratum: [Phys.\ Rev.\ D {\bf 60} (1999) 049902]
  doi:10.1103/PhysRevD.60.049902, 10.1103/PhysRevD.58.021302
  [hep-ph/9709417].



\bibitem{rajantie2}
  M.~Herranen, T.~Markkanen, S.~Nurmi and A.~Rajantie,
  Phys.\ Rev.\ Lett.\  {\bf 115} (2015) 241301
  doi:10.1103/PhysRevLett.115.241301
  [arXiv:1506.04065 [hep-ph]].
  
\bibitem{ema}
  Y.~Ema, K.~Mukaida and K.~Nakayama,
  JCAP {\bf 1610} (2016) no.10,  043
  doi:10.1088/1475-7516/2016/10/043
  [arXiv:1602.00483 [hep-ph]].


\bibitem{Kohri:2016wof}
  K.~Kohri and H.~Matsui,
  Phys.\ Rev.\ D {\bf 94} (2016) no.10,  103509
  doi:10.1103/PhysRevD.94.103509
  [arXiv:1602.02100 [hep-ph]].


\bibitem{Kohri:2016qqv}
  K.~Kohri and H.~Matsui,
  arXiv:1607.08133 [hep-ph].


\bibitem{Enqvist:2016mqj}
  K.~Enqvist, M.~Karciauskas, O.~Lebedev, S.~Rusak and M.~Zatta,
  JCAP {\bf 1611} (2016) 025
  doi:10.1088/1475-7516/2016/11/025
  [arXiv:1608.08848 [hep-ph]].

\bibitem{BirrellDavies}
  N.~D.~Birrell and P.~C.~W.~Davies,
  doi:10.1017/CBO9780511622632

\bibitem{Keldysh:1964ud}
  L.~V.~Keldysh,
  Zh.\ Eksp.\ Teor.\ Fiz.\  {\bf 47} (1964) 1515
   [Sov.\ Phys.\ JETP {\bf 20} (1965) 1018].

\bibitem{Schwinger:1960qe}
  J.~S.~Schwinger,
  J.\ Math.\ Phys.\  {\bf 2} (1961) 407.
  doi:10.1063/1.1703727.
\bibitem{Jordan:1986ug}
  R.~D.~Jordan,
  Phys.\ Rev.\ D {\bf 33} (1986) 444.
  doi:10.1103/PhysRevD.33.444
\bibitem{Calzetta:1986ey}
  E.~Calzetta and B.~L.~Hu,
  Phys.\ Rev.\ D {\bf 35} (1987) 495.
  doi:10.1103/PhysRevD.35.495
\bibitem{Calzetta:1986cq}
  E.~Calzetta and B.~L.~Hu,
  Phys.\ Rev.\ D {\bf 37} (1988) 2878.
  doi:10.1103/PhysRevD.37.2878
 


\bibitem{Ringwald:1986wf}
  A.~Ringwald,
  Z.\ Phys.\ C {\bf 34} (1987) 481.
 doi:10.1007/BF01679866

\bibitem{Ringwald:1987ui}
  A.~Ringwald,
  Annals Phys.\  {\bf 177} (1987) 129.
  doi:10.1016/S0003-4916(87)80027-1

\bibitem{Baacke:1997rs}
  J.~Baacke, K.~Heitmann and C.~Patzold,
  Phys.\ Rev.\ D {\bf 56} (1997) 6556
  doi:10.1103/PhysRevD.56.6556
  [hep-ph/9706274].

\bibitem{Baacke:1996se}
  J.~Baacke, K.~Heitmann and C.~Patzold,
  Phys.\ Rev.\ D {\bf 55} (1997) 2320
  doi:10.1103/PhysRevD.55.2320
  [hep-th/9608006].



\bibitem{Parker:1974qw}
  L.~Parker and S.~A.~Fulling,
  Phys.\ Rev.\ D {\bf 9} (1974) 341.
  doi:10.1103/PhysRevD.9.341

\bibitem{Fulling:1974pu}
  S.~A.~Fulling, L.~Parker and B.~L.~Hu,
  Phys.\ Rev.\ D {\bf 10} (1974) 3905.
  doi:10.1103/PhysRevD.10.3905

  \bibitem{Bunch:1980vc}
  T.~S.~Bunch,
  J.\ Phys.\ A {\bf 13} (1980) 1297.
  doi:10.1088/0305-4470/13/4/022

\bibitem{tranberg}
  T.~Markkanen and A.~Tranberg,
  JCAP {\bf 1308} (2013) 045
  doi:10.1088/1475-7516/2013/08/045
  [arXiv:1303.0180 [hep-th]].

  \bibitem{Paz:1988mt}
  J.~P.~Paz and F.~D.~Mazzitelli,
  Phys.\ Rev.\ D {\bf 37} (1988) 2170.
  doi:10.1103/PhysRevD.37.2170



\bibitem{Winitzki}
  V.~Mukhanov and S.~Winitzki,


  \bibitem{Coleman}
  S.~R.~Coleman and E.~J.~Weinberg,
  Phys.\ Rev.\ D {\bf 7} (1973) 1888.
  doi:10.1103/PhysRevD.7.1888


\bibitem{moss}
  I.~G.~Moss,
  arXiv:1509.03554 [hep-th].

  \end{thebibliography}
\end{document}